\documentclass[colorlinks,citecolor=blue,urlcolor=blue, linkcolor=blue]{article}

\usepackage{arxiv}

\usepackage[utf8]{inputenc} 
\usepackage[T1]{fontenc}    
\usepackage{url}            
\usepackage{booktabs}       
\usepackage{amsfonts}       
\usepackage{nicefrac}       
\usepackage{microtype}      
\usepackage{graphicx}
\usepackage{natbib}
\usepackage{doi}

\usepackage{float}
\usepackage{subcaption}
\usepackage{xcolor} 
\usepackage{hyperref} 

\usepackage{secdot}
\usepackage{array,epsfig,rotating}
\usepackage{amsmath}
\usepackage{amssymb}
\usepackage{amsfonts}
\usepackage{multirow}
\usepackage{amsthm}
\usepackage{bm}
\usepackage{subcaption}
\usepackage{tabularx}
\usepackage{longtable}
\usepackage{mathrsfs}

\usepackage{cleveref}       

\usepackage{ulem}

\def\bSig\mathbf{\Sigma}

\def \bone {\mbox{\boldmath $1$}}

\AtBeginDocument{%
  \expandafter\def\expandafter\@tabarray\expandafter{%
    \expandafter\CT@start\@tabarray}}

\title{Restricted Mean Survival Time Estimation Using Bayesian Nonparametric Dependent Mixture Models}

\author{ \href{https://www.hopkinsmedicine.org/profiles/details/ruizhe-chen}{
Ruizhe Chen}\thanks{Corresponding Author; ORCID: https://orcid.org/0000-0003-3924-3328} \\
	Division of Quantitative Sciences\\
	The Sidney Kimmel Comprehensive Cancer Center\\
        School of Medicine\\
        Johns Hopkins University\\
	Baltimore, MD 21205 \\
	\texttt{rchen89@jhmi.edu} \\
	\And
	\href{https://publichealth.uic.edu/profiles/sanjib-basu/}{
 Sanjib Basu} \\
	Division of Epidemiology and Biostatistics\\
	School of Public Health\\
	University of Illinois Chicago\\
        Chicago, IL 60612\\
	\texttt{sbasu@uic.edu} \\
	\AND
	\href{https://www.mayo.edu/research/faculty/shi-qian-ph-d/bio-00096527}{Qian Shi} \\
	Division of Clinical Trials and Biostatistics \\
        Department of Quantitative Health Sciences \\
        Mayo Clinic \\
	Rochester, MN 55905 \\
	\texttt{shi.qian2@mayo.edu}
}

\renewcommand{\shorttitle}{RMST Estimation Using BNP Dependent Mixture Models}

\hypersetup{
pdftitle={Restricted Mean Survival Time Estimation Using Bayesian Nonparametric Dependent Mixture Models},
pdfsubject={biostatistics},
pdfauthor={Ruizhe Chen, Sanjib Basu, Qian Shi},
pdfkeywords={Restricted Mean Survival Time \and Bayesian Non-parametric Inference \and Causal Inference},
}

\begin{document}
\maketitle

\begin{abstract}
	Restricted mean survival time (RMST) is an intuitive summary statistic for time-to-event random variables, and can be used for measuring treatment effects. Compared to hazard ratio, its estimation procedure is robust against the non-proportional hazards assumption. We propose nonparametric Bayeisan (BNP) estimators for RMST using a dependent stick-breaking process prior mixture model that adjusts for mixed-type covariates. The proposed Bayesian estimators can yield both group-level causal estimate and subject-level predictions. Besides, we propose a novel dependent stick-breaking process prior that on average results in narrower credible intervals while maintaining similar coverage probability compared to a dependent probit stick-breaking process prior. We conduct simulation studies to investigate the performance of the proposed BNP RMST estimators compared to existing frequentist approaches and under different Bayesian modeling choices. The proposed framework is applied to estimate the treatment effect of an immuno therapy among KRAS wild-type colorectal cancer patients.
\end{abstract}


\section{Introduction \label{sec:intro}}

In clinical research, one is often interested in quantifying covariate effect on a time-to-event, often observed with potential censoring. The Cox proportional-hazards model and the resulting hazard ratios (HRs) are the ``go-to'' approach for such analysis. However, interpreting an HR becomes difficult in presence of non-proportional hazards, which can occur when, for example, time-varying covariate effects exist. In recent years, the application of restricted mean survival time (RMST) in planning and analyzing randomized clinical trials (RCTs) with time-to-event endpoints have drawn the attention of many in the field of medical/clinical statistics \citep{tian2018efficiency,freidlin2021restricted,royston2011use,zhang2012double,uno2014moving,uno2015alternatives,weir2021multivariate,royston2013restricted,wei2015meta,tian2020empirical}. The popularity of RMST can be attributed to the potential benefits from using it as a measure of treatment effect in survival analysis over other conventional measures such as the HR. The interpretation of RMST is intuitive, clinically relevant, and is model-free in the sense that may not rely on assumptions such proportional hazards. Besides, RMST summarizes survival over a fixed follow-up time period and is of inherent interest in settings where cumulative covariate effects are appealing \citep{wang2018modeling}. 

There is an abundance of frequentist methods for estimating RMST and its associated variances in the literature. In an ideal RCT setting, RMST can be estimated consistently by the area under the Kaplan-Meier (KM) curve up to a specific  time $\tau$ given that  $\tau$ is less than or equal to the maximum observed event time and under non-informative censoring \citep{klein2003survival,tian2014predicting}. 
In addition to the nonparametric RMST estimators introduced by \cite{irwin1949standard} and \cite{meier1975estimation}, researchers have proposed various RMST regression methods. As summarized by \cite{wang2018modeling}, these approaches in general, estimate the regression parameters and baseline hazard from a Cox model, calculate the cumulative baseline hazard, which are transformed to obtain the survival function and, and integrate the survival function to obtain the RMST. Another category of RMST modeling approaches resembles the accelerated failure time model by assuming a linear relationship between covariates and $E[log(T)]$ as the response variable \citep{tian2014predicting,ambrogi2022analyzing}. There are only a few works on Bayesian inference in RMST. \cite{poynor2019nonparametric} studied a related yet different problem of Bayesian inference for the mean residual life (MRL) function, defined as the expected remaining survival time given survival up to time $\tau$. They developed a nonparametric Bayesian (BNP) inference approach for MRL functions by constructing a Dirichlet process (DP) mixture model for the underlying survival distribution. \cite{zhang2022bayesian} proposed BNP estimators for RMST, for both right and interval censored data, assuming mixture of Dirichlet process priors. 

In this article, we develop a Bayesian nonparametric dependent mixture (BNPDM) approach for regression modeling in RMST. We utilize these models to make inference about individual-level RMST difference (RMSTD), as well as population-level causal average treatment effect (ATE). We explore different prior choices for a dependent stick-breaking process (DSBP) mixture model, which includes: ($i$) A finite-dimensional predictor-dependent stick-breaking prior via sequential logistic regressions \citep{rigon2021tractable,ishwaran2001gibbs}; ($ii$) the dependent probit-stick breaking process prior \citep{rodriguez2011nonparametric}; ($iii$) our proposed novel shrinkage probit-stick breaking process prior, which is a data-adaptive stick-breaking prior based on a probit regression model. 
Research interest in RMST inference is often centered around comparing group differential RMST at one or multiple $\tau$'s as a fixed-time analysis. However, only looking at RMST values at a single time point may not accurately reflect the totality of clinical effect and may be misleading about clinical significance of the experimental treatment \citep{freidlin2019methods}. We provide point-wise Bayesian estimate and inference for the entire RMST curve. 
We evaluate and compare performance of the proposed BNPDM models with two existing non-Bayesian RMST approaches \citep{tian2014predicting,ambrogi2022analyzing} through extensive simulation studies. 

The rest of this paper is organized as follows. In \ref{RMST-BNPDP-chap-method}, we introduce our proposed BNPDM models for drawing RMST inference. In \ref{BNP-gSB-chap-sim-study}, we conduct simulation studies to examine the performance of the proposed BNPDM models both under different prior choices and compared to existing frequentist methods \citep{tian2014predicting,ambrogi2022analyzing}. In \ref{BNP-gSB-chap-real-data}, we present an application of our proposed BNPDM models to analyze real data from a phase III colorectal cancer trial. In \ref{BNP-gSB-chap-discuss-future}, we summarize our findings and give our thoughts on the characteristics of the proposed estimators.

\section{Methodology \label{RMST-BNPDP-chap-method}}
\subsection{\label{RMST-BNPDP-chap-method-gSB} Nonparametric Bayesian Inference of Restricted Mean Survival Time}

Let $T$ denote a random variable with non-negative support representing time from an appropriate time origin to a clinical event of interest, and assume that $T$ is subject to non-informative right censorship due to either a random drop-out or reaching a maximum follow-up time. 
RMST is defined as $\mu(\tau)= E[\min(T,\tau)] = \int_0^{\tau}S(t)dt$ where $S(\cdot)$ denotes the survival function of $T$. Thus, RMST can be interpreted as the average of all potential event times measured (from time $0$) up to $\tau$ and mathematically measured as the area under the survival curve up to $\tau$.

For predictor $\bm{w}$, we model the density  of  $T$ as  predictor-dependent mixtures of a predictor-dependent general kernel density as
\begin{align}
\label{DSBP-construct-def0}
    f(t \; | \; G_{\bm{w}}) = \int K_{\bm{w}}(t \; | \; \bm{\theta}) \; dG_{\bm{w}}(\bm{\theta}) = \sum_{h=1}^L \pi_h(\bm{w}) K_{\bm{w}}(t \; | \; \bm{\theta}_h); \qquad t \in \mathbb{R}^+
\end{align}
where 
\begin{align}
\begin{split}
\label{DSBP-construct-def1}
    &G_{\bm{w}}(\cdot) = \sum_{h=1}^{L} \Bigg\{v_h(\bm{w}) \prod_{l < h}\big[1-v_l(\bm{w})\big]\Bigg\} \delta_{\bm{\theta}_h(\bm{w})}(\cdot) ; \quad 1 \leq L \leq \infty. 
\end{split}
\end{align}
Here 
$\delta_{\bm{\theta}_h(\bm{w})}(\cdot)$ is the Dirac measure at $\bm{\theta}_h(\bm{w})$, $\pi_h(\bm{w}) = v_h(\bm{w}) \prod_{l < h}\big[1-v_l(\bm{w})\big]$, and $\pi_h(\bm{w}) \geq 0$ are random functions of $\bm{w}$ such that $\sum_{h=1}^L \pi_h(\bm{w})=1$ a.s. for each fixed $\bm{w} \in \mathscr{W}$. $v_h(\bm{w}), h \in \mathbb{N}$, are $[0,1]$-valued (predictor-dependent) stochastic processes, independent from $\bm{\theta}_h(\bm{w})$, with index set $\mathscr{W}$. $v_h(\bm{w})$ can be viewed as a transition kernel such that for all $w \in \mathscr{W}$, $v_h(w,\mathscr{A})$ is a probability measure, and for all $\mathscr{A} \in \mathscr{B}(\mathscr{W})$ (a Borel $\sigma$-field on $\mathscr{W}$), $v_h(w,\mathscr{A})$ is measurable. There is some flexibility for choosing an appropriate transition kernel. For example, \cite{dunson2008kernel} chose $v_h(\bm{w}) = v_h K(\bm{w}, \Gamma_h), v_h \sim Beta(1,\lambda)$, and $\Gamma \sim H$ is a location where $K(\cdot): \mathbb{R} \times \mathbb{R} \rightarrow [0,1]$ is a bounded kernel function. We propose the following formulation for the $h^{th}$ stick-breaking probability:
\begin{align}
\label{DSBP-construct-def2}
    v_h(\bm{w}) = g(\bm{\psi}(\bm{w})^{\prime} \bm{\alpha}_h), \quad \bm{\alpha}_h \sim Q
\end{align}
for a link function $g(\cdot): \mathbb{R} \rightarrow [0,1]$, $\bm{\psi}(\bm{w})=(\psi_1(w_1),\ldots,\psi_R(w_R))$ denoting $R$ functions of covariate $\bm{w}=\{w_1,\ldots,w_R\}$ and random measure $Q$ is defined on $\mathbb{R}^R$. Subsequently, a predictor-dependent stick-breaking process can be defined as
\begin{align}
\begin{split}
\label{PDSBP-construct-def1}
    &\pi_1(\bm{w}) = v_1(\bm{w}) \\  &\pi_h(\bm{w})=\big(1-v_1(\bm{w})\big)\big(1-v_2(\bm{w})\big)\ldots\big(1-v_{h-1}(\bm{w})\big)v_h(\bm{w}), \ h=2,\ldots,L-1. \\
    &\pi_L(\bm{w})=1-\sum_{h=1}^{L-1} \pi_h(\bm{w}) = \big(1-v_1(\bm{w})\big)\ldots\big(1-v_{L-1}(\bm{w})\big).
\end{split}
\end{align}

For a finite $L$, the construction of the weights in (\ref{PDSBP-construct-def1}) ensures that $\sum_{h=1}^L v_h=1$. By applying the linear form $\bm{\psi}(\bm{w})^\prime \bm{\alpha}_h,$ for certain function $\bm{\psi}(\cdot)$, one can include mixed-type (both continuous and discrete) predictors $\{w_r\}$ such that $v_h(\bm{w}): \mathbb{R} \times,\cdots,\times \{0,1,\ldots,K_r\} \times \cdots~ \rightarrow [0,1]$ where, for example, a discrete values $\bm{w}_r$ may have support on $\{0,1,\ldots,K_r\}$. The link function $g(\cdot)$ can be chosen as an inverse logit link as for the case of logistic stick-breaking process priors \cite{ren2011logistic} and logit stick-breaking process priors \citep{rigon2021tractable}, or a probit link as for the case of probit stick-breaking process priors \citep{chung2009nonparametric,rodriguez2011nonparametric,pati2014bayesian}. Similarly, we model the kernel density function with predictor-dependent parameterizations. Suppose that $K \big(\cdot | \; \bm{\theta})$  with $\bm{\theta}= (\eta, \omega)$ is a two-parameter kernel density, for example, a  lognormal density with parameters $\eta$ and  $\omega$, or 
a Weibull density with scale $\eta$ and shape $\omega$, or a (two-parameter) Gamma density with rate $\eta$ and shape $\omega$. We model the $h^{th}$ cluster kernel density by incorporating predictor dependence on $\eta$ such that
\begin{align}
\label{DSBP-construct-kernel-def-1}
    K_{\bm{w}}(t \; | \; \bm{\theta}_h) = K\big(t \; | \; \eta_h(\bm{w}) = \bm{\psi}(\bm{w})^T \bm{\beta}_h, \omega_h \big)
\end{align}
where $\eta_h(\bm{w})$ is defined as a linear combination of $\bm{\psi}(\bm{w})^T$ and atoms $\bm{\beta}_h$. Therefore, the atom sampling process in (\ref{DSBP-construct-def1}) is given by
\begin{align}
\label{DSBP-construct-kernel-def-2}
    \bm{\theta}_h(\bm{w})=\big(\bm{\psi}(\bm{w})^T \bm{\beta}_h, \omega_h\big), \quad \bm{\theta}_h=(\bm{\beta}_h, \omega_h), \quad \bm{\theta}_h \sim P
\end{align}
for some random measure $P$.

Under the above formulations (\ref{DSBP-construct-def0})-(\ref{DSBP-construct-def2}), the survival function can also be represented in a constructive form as:
\begin{align*}
    \begin{split}
        S(t \; | \; G_{\bm{w}}) &= \int_t^{\infty} f(t \; | \; G_{\bm{w}}) dt = \int_t^{\infty} \sum_{h=1}^L \pi_h(\bm{w}) K_{\bm{w}}\big(t \; | \; \bm{\theta}_h\big) dt \\
        &= \sum_{h=1}^L \pi_h(\bm{w}) \int_t^{\infty} K_{\bm{w}}\big(t \; | \; \bm{\theta}_h\big) dt =  \sum_{h=1}^L  \pi_h(\bm{w}) S_{\bm{w}}\big(t \; | \; \bm{\theta}_h\big)
    \end{split}
\end{align*}

Similarly for the RMST function,
\begin{align}
\label{dependent-mix-eq-2}
    RMST\big(t \; | \; G_{\bm{w}}\big) = \sum_{h=1}^\infty \pi_h (\bm{w}) RMST_{ \bm{w}}\big(t \; | \;\bm{\theta}_h\big)
\end{align}

We show, in the supplemental materials, that the kernel RMST function is analytically tractable if the kernel density $K(t \; | \;\cdot)$ assumes a Weibull or Gamma form. These convenient structures allow us to formulate individual and group-level BNP estimators in closed form expressions and result in improved computational efficiency. 
Alternatively, one can assume predictor dependence imposes only on either the mixing probabilities or the kernel density which defines a single-atoms predictor-dependent stick-breaking process mixture model pr a single-$\pi$ linear dependent Dirichlet process (LDDP) mixture model, respectively.




\subsection{Shrinkage Probit Stick-Breaking Process Prior  \label{sec-SPSBP-prior-and-related}}


We propose a novel DSBP prior that is inspired by \citep{rodriguez2011nonparametric,ren2011logistic,rigon2021tractable}. Given a covariate matrix of $N$ observations from $R$ covariates $\bm{W}=\{w_{1,1},\ldots,w_{R,N}\}$ of both continuous and discrete type, suppose we consider a stick-breaking probability assignment mechanism defined by (\ref{DSBP-construct-def2}) where a multivariate normal prior is assumed for $Q$, namely,
    $\bm{\alpha}_h=(\alpha_{h,1},\ldots,\alpha_{h,R}) \sim \mathcal{N}_{R}\big(\bm{\mu}_{\alpha}, \bm{\sigma^2}_{\alpha} \bm{I}_R\big)$ 
where $(\bm{\mu}_{\bm{\alpha}}, \bm{\sigma}_{\alpha})$ is specified \textit{a priori}. 
For this choice of $Q$, we have 
\begin{align}
        \label{SPSBP-prior-def2}
    \psi(\bm{w}_i)^{\prime} \bm{\alpha}_h 
   \sim \mathcal{N}_1\left(\sum_{r=1}^R  \psi_r(w_{ir}) \mu_{\alpha_r}.~
  \sum\limits_{r=1}^R \psi_r(w_{ir})^2 
   \sigma_{\alpha_{r}}^2 \right),~h = 1,\ldots,L-1.
%
\end{align}


We propose a DSBP prior based on the above form 
in which the stick-breaking probability of the $i^{th}$ observation for the $h^{th}$ cluster is modeled by a probit link as
\begin{equation}
\label{SPSBP-prior-def1}
    v_h(\bm{w}_i) = g\big(\bm{\psi}(\bm{w}_i)^{\prime} \bm{\alpha}_h\big) = \Phi\Big(\frac{\bm{\psi}(\bm{w}_i)^{\prime} \bm{\alpha}_h - \bm{\mu}(\bm{W},\bm{\alpha}_h)}{\sigma(\bm{W},\bm{\alpha}_h)}\Big)
\end{equation}

where $\Phi(\cdot)$ is the CDF of a standard normal distribution, and the location and scale functions are specified as 
\begin{align}
\label{SPSBP-prior-def2}
     \mu(\bm{W},\alpha_h) = \frac{1}{N} \sum_{i=1}^N \bm{\psi}(\bm{w}_{i})^{\prime} \bm{\alpha}_{h}, ~\hbox{and}~
     \sigma^2(\bm{W},\alpha_h) = \sum\limits_{i=1}^N \Big( \bm{\psi}(\bm{w})_i^{\prime} \bm{\alpha}_h - \mu(\bm{W},\alpha_h)\Big)^2/(N-1) 
\end{align}
We refer to the model in (\ref{SPSBP-prior-def1}) based on the link $g(s) = \Phi\big((s - \bm{\mu}(\bm{W},\alpha_h))/\sigma(\bm{W},\alpha_h)\big)$ 
as shrinkage probit model
. The linear transformation function 
$\bm{\psi}(\bm{w}_i)^{\prime} \bm{\alpha}_h$) is flexible and can accommodate mixed-type predictors. The SPSBP prior is distinct from (dependent) PSBP prior as the latter  would assign stick breaking probabilities as $v_h(\bm{w}_i) = 
\Phi\big(\bm{\psi}(\bm{w}_i)^\prime \bm{\alpha}_h \big)$
The SPSBP prior 
is built on the basic structure of the (dependent) PSBP prior, yet it
adds a feature of borrowing information from clusterings of predictors when assigning stick-breaking probabilities, an idea that was implemented using kernel functions in \citep{dunson2007bayesian,dunson2008kernel,ren2011logistic}. Consider a sample predictor matrix $\bm{W}_{N \times R}$ and its linearly transformed components at the $h^{th}$ cluster: $\psi(\bm{W})^{\prime}\alpha_h = \{\psi(\bm{w}_i)^{\prime} \bm{\alpha}_h, i=1,\ldots,N\}$ where $\bm{\alpha}_h \sim N_R(\mu,\Sigma), h=\{1,\ldots,L-1\}$. Instead of assigning stick-breaking probabilities ($\pi_{ih}$) for $\psi(\bm{w}_i)^\prime \bm{\alpha}_h$ based on its location in the empirical cumulative distribution function of $\psi(\bm{W}^\prime)\alpha_h$ with a standard normal probit model, the SPSBP analogously assigns $\pi_{ih}$ according to a probit model with mean and variance set equal to the sample first and second central moments of $\psi(\bm{W})^\prime\alpha_h$, i.e., the shrinkage probit model. 

\begin{figure}[H]
\centering
 \renewcommand{\baselinestretch}{1}
\caption[SPSBP versus PSPB prior]{SPSBP versus PSPB prior: differences between with shrinkage effect and without shrinkage effect in assigning stick-breaking probabilities}
     \begin{subfigure}[b]{\textwidth}
     \includegraphics[width=\textwidth]{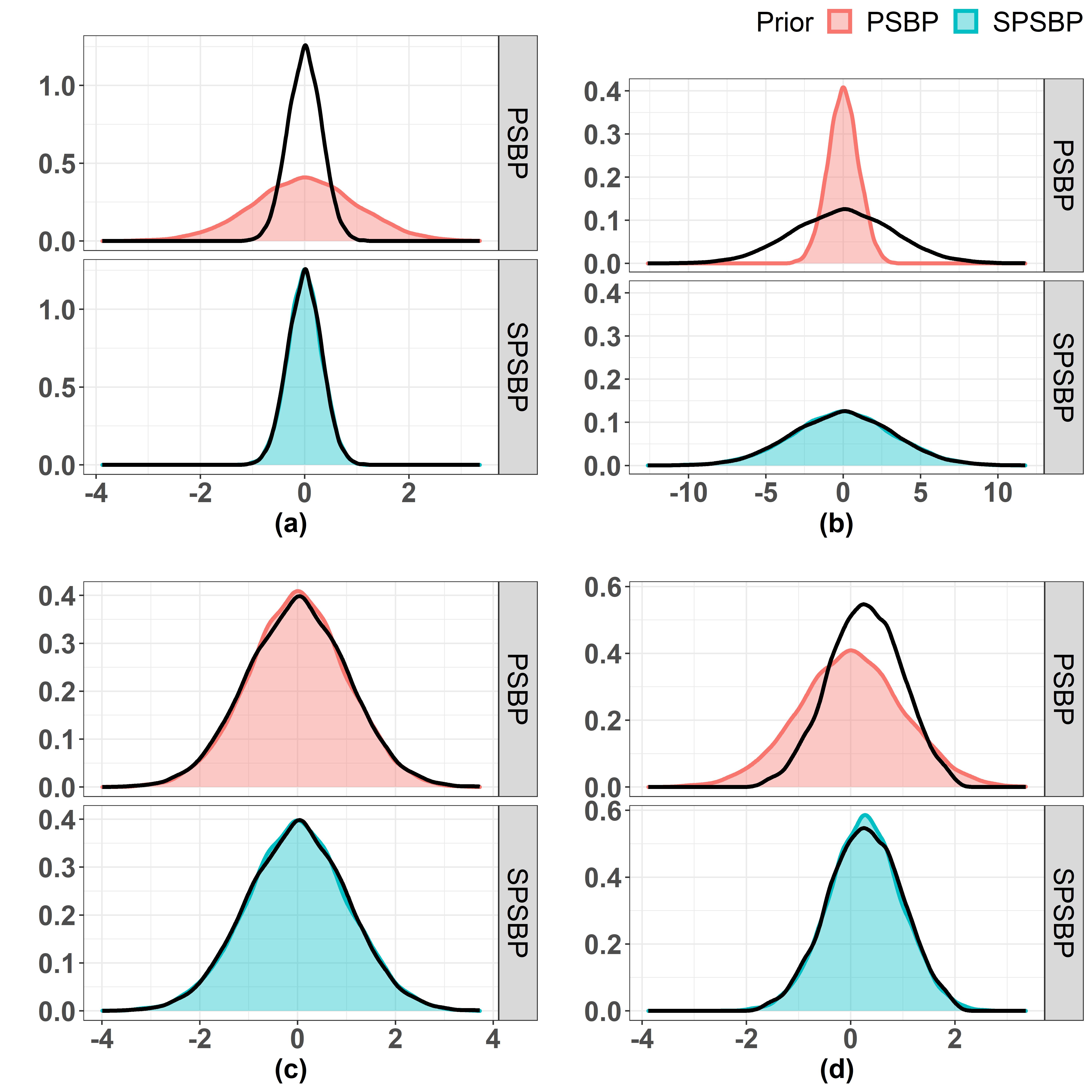}
     \end{subfigure}
    \caption*{Density plots for sample observations from $\psi(\bm{W})^\prime\alpha_h = \{\psi(\bm{w}_i)^\prime \bm{\alpha}_h, i=1,\ldots,N\}$ (black solid lines), a standard normal (PSBP prior), and a normal distribution with $\mu$ and $\sigma$ defined by \ref{SPSBP-prior-def2} (SPSBP prior); $\psi(\bm{w}_i)^\prime \bm{\alpha}_h$ is defined in \ref{SPSBP-prior-def2}; four $\bm{W}$ generation scenarios ($N=10,000$) are considered: ($a$) $\bm{W} \sim \mathcal{N}_4(\bm{0},0.1 \cdot I_4)$, ($b$) $\bm{W} \sim \mathcal{N}_4(\bm{0},10 \cdot I_4)$, ($c$) $\bm{W} \sim \mathcal{N}_4(\bm{0}, I_4)$, (d) same predictors generation scheme (RCT setting) as specified in the simulation studies section: $\bm{W}=\{A,X_1,X_2,X_3\}, A \sim Binomial(10,000,0.5), X_1 \sim N(0,1), X_2 \sim Binomial(10,000,0.7), X_3 \sim t(d.f.=5,ncp=0)$.}
    \label{fig-chapter-3-0}
\end{figure}

The SPSBP prior utilizes the clustering information of $\psi(\bm{W})^\prime\alpha_h$ around $\psi(\bm{w}_i)^\prime \bm{\alpha}_h$, thus shrinking the difference between the empirical distribution of $\psi(\bm{W})^\prime\alpha_h$ and the probit model that assigns $\pi_{ih}$ (for $\psi(\bm{w}_i)^\prime \bm{\alpha}_h$). On the other hand, no such shrinkage effect exists for the (dependent) PSBP prior since $\pi_{ih}$ are assigned uniformly based on $\Phi(\cdot)$. Figure \ref{fig-chapter-3-0} shows density plots for sample observations from $\psi(\bm{W})^\prime\alpha_h = \{\psi(\bm{w}_i)^\prime \bm{\alpha}_h, i=1,\ldots,N\}$ (black solid lines), a standard normal (PSBP prior), and a normal distribution with mean and variance set according to (\ref{SPSBP-prior-def2}) (SPSBP prior). Linearly transformed component $\psi(\bm{w}_i)^\prime \bm{\alpha}_h$ is defined in (\ref{SPSBP-prior-def2}). Four covariates matrix ($\bm{W}$) generation scenarios (sample size $N=10,000$) are considered: ($a$) $\bm{W} \sim \mathcal{N}_4(\bm{0},0.1 \cdot I_4)$; ($b$) $\bm{W} \sim \mathcal{N}_4(\bm{0},10 \cdot I_4)$; ($c$) $\bm{W} \sim \mathcal{N}_4(\bm{0}, I_4)$; (d) same predictors generation scheme (RCT setting) as specified in the simulation studies section: $\bm{W}=\{A,X_1,X_2,X_3\}, A \sim Binomial(10,000,0.5), X_1 \sim N(0,1), X_2 \sim Binomial(10,000,0.7)$, $X_3 \sim t(d.f.=5)$. As shown by Figure \ref{fig-chapter-3-0}, this shrinkage effect persists under various scenarios, e.g, when the scale of $\psi(\bm{W})^\prime\alpha_h$ is either smaller, equal to, or larger than that of a standard normal, or in presence of a shifted location. Measuring the distance between $\psi(\bm{w}_i)^\prime \bm{\alpha}_h$ and $\psi(\bm{w}_j)^\prime \bm{\alpha}_h$ (where $i\neq j$) is realized through linear combinations and the multivariate Gaussian prior assumption on $\alpha_h$, which is motivated to accommodate mixed-type (both discrete and continuous) predictors. In replicated simulation studies, we observe that this shrinkage effect brings the benefit of obtaining `shrunk credible intervals and smaller RMSEs, when estimating group-level ATE measured by RMST difference (RMSTD) compared to the PSBP prior. 
Therefore, the SPSBP prior (\ref{SPSBP-prior-def1}) approximates a DP with precision parameter $1$ marginally for each fixed $w \in \mathcal{W}$ given a sufficiently large $N$. 
In our application of the SPSBP prior for RMST inference, we assume that $\bm{\alpha}_h \sim \mathcal{N}_{R}\big(\bm{0}, \bm{\sigma}^2_{\alpha} \bm{I}_R\big)$, which leads to $\mu_{\alpha_1}+\ldots +\mu_{\alpha_R}=0$. In this setting we can also conveniently apply $v_h(\bm{w}_i) = \Phi\Big(\frac{\bm{w}^\prime_i \bm{\alpha}_h}{\sigma_{v_h}(\bm{W},\bm{\alpha}_h)}\Big)$. The PSBP prior in \cite{papageorgiou2015bayesian} for modeling spatially indexed data of mixed type followed a similar structure.
Their model allows for observations that correspond to nearby areas to be more likely to have similar values for the component weights than observations from areas that are far apart. In their formulation, $\alpha_h(\bm{w}_i)$ are realizations of marginal Gaussian Markov random fields 
and the level of borrowing on clustering of covariates is controlled by pre-specified parameters of the random field. In comparison, the level of shrinkage  in  our proposed method in (\ref{SPSBP-prior-def2}) is mostly data-dependent.

Another DSBP prior that can adjust for mixed-type predictors is the logit stick-breaking process (LSBP) prior introduced by  \cite{rigon2021tractable}. Assuming a LSBP prior, the $h^{th}$ stick-breaking probability is given by
\begin{equation*}
    v_h(\bm{w}_i)=1/\{1-\exp(\bm{\psi}(\bm{w}_i)^\prime \bm{\alpha}_h) \}, \quad \bm{\alpha}_h \sim \mathcal{N}_R(\bm{\mu},\bm{\Sigma}), \ h \in \{1,\ldots,L-1\} 
\end{equation*}

\subsection{A Bayesian Non-parametric Inference Framework}

\subsubsection{Subject-Level and Group-Level Estimands and Estimators}

In \ref{RMST-BNPDP-chap-method-gSB}, we provided a general definition of RMST as the expected survival time of time-to-event $T$ restricted to certain time point $[0,\tau]$. Conditional on the predictor matrix $\bm{W}$ and a fixed time point $\tau$, a conditional RMST function can be defined as
\begin{align}
\label{subj-estimand-def-1}
    RMST_{\bm{\theta}}(t=\tau \; | \; \bm{W}) = \int_0^\tau S_{\bm{\theta}}(t \; | \; \bm{W}) \; dt,
\end{align}


which relates to the (marginal) RMST function by
\begin{align}
\label{subj-estimand-def-2}
   RMST(t=\tau) = E_{\bm{W}}\big[RMST(t=\tau \; | \; \bm{W})\big] = \int_{-\infty}^\infty \int_0^\tau S_{\bm{\theta}}(t \; | \; \bm{W}) f_{\bm{W}}(\bm{w}) \; dt d\bm{W}
\end{align}

\cite{tian2014predicting} studied a class of frequentist regression models for estimating the conditional RMST 
given  ``baseline'' covariate. In this section, we develop nonparametric Bayes conditional RMST inference,  
Let the observed $i^{th}$ time-to-event be $(Y_i,\delta_i)$ where $Y_i = \text{min}(T_i,C_i)$, $\delta_i = I(T_i \leq C_i)$ and $C_i$ is the censoring variable including a maximum follow-up time of $\tau_c$. Also, $\bm{W}_i = (A_i,\bm{X}_i)$ where $\bm{X}_i$ is a vector of time-fixed covariate and $A_i$ is binary treatment group indicator. With a slight abuse of notation, let $\bm{\theta}$ denote the combined parameter vector of both the stick-breaking process prior and the kernel density.
The posterior mean conditional RMST function for the $i^{th}$ subject is 
\begin{align}
\label{subj-estimator-def-1}
    \widehat{RMST}(t=\tau \; | \; \bm{w}_i) = E_{\bm{\theta}} \left[\int_0^\tau S_{\bm{\theta}}(s \; | \; \bm{w}_i) \; ds \;\;|\bm{Y},\bm{\delta} \right ]
\end{align}
which can be estimated based on Markov chain samples $\{\bm{\theta}_1,\ldots,\bm{\theta}_L\}$  and based on our 
 model constructions in (\ref{DSBP-construct-def0}--\ref{dependent-mix-eq-2}) as 

\begin{align}
\label{subj-estimator-def-2}
    \widehat{RMST}(t = \tau \; | \; \bm{w}_i) = E_{\bm{\theta}}\Bigg[\sum_{h=1}^L \Big[ \pi_h\big(\bm{\psi}_1(\bm{w}_i)^\prime\bm{\alpha}_h\big) \int_0^\tau \int_t^\infty K\big(s \; | \; \bm{\psi}_2(\bm{w}_i)^\prime\bm{\beta}_h,\bm{\omega}_h\big) \; ds dt \Big]\Bigg]
\end{align}




where $1\leq L \leq \infty$, $\bm{\theta}=\{\bm{\alpha},\bm{\beta},\bm{\omega}\}$. Here we distinguish the linear transformation function ($\psi_1(\cdot)$) applied in the stick-breaking probabilities from the one ($\psi_2(\cdot)$) applied in the kernel density with different subscripts. Consequently, a $100(1-\alpha)\%$ credible interval (CI) for $RMST$ is $[\widehat{RMST}^{(\alpha/2)},\widehat{RMST}^{(1-\alpha/2)}]$ where $\widehat{RMST}^{(\alpha/2)}$ and $\widehat{RMST}^{(1-\alpha/2)}$ are calculated as the $(\alpha/2)$th and $(1-\alpha/2)$th quantiles of the posterior distribution of $RMST$.



\cite{chen2001causal} defined an average (marginal causal) treatment effect (ATE) in terms of the average group differential RMSTD under the counter-factual framework \citep{morgan2015counterfactuals} by, that is,
\begin{align}
\label{causal-estimand-def-1}
\begin{split}
    \Delta &= RMST_{A_1}(t=\tau) - RMST_{A_0}(t=\tau) \\ &= \int_0^\tau E_{\bm{X}}\Big[S_{\bm{\theta}}(t \; | \;A=1, \bm{X})\Big] -  E_{\bm{X}}\Big[S_{\bm{\theta}}(t \; | \;A=0, \bm{X})\Big] \; dt 
\end{split}
\end{align}

We define marginal Bayesian estimators of the causal estimand $\Delta$ defined in (\ref{causal-estimand-def-1}) using the empirical distribution of $\bm{X}$ as a nonparametric estimator of $f_{\bm{X}}(\bm{x})$
\begin{align}
    \begin{split}
    \label{causal-estimator-BNP-def-1}
        \widehat{\Delta} 
        &= \frac{1}{N} \sum_{i=1}^N 
        E_{\bm{\theta}} \left[\int_0^\tau S_{\bm{\theta}}(t \; | \;A=1, \bm{X}_i=\bm{x}_i) - S_{\bm{\theta}}(t \; | \;A=0, \bm{X}_i=\bm{x}_i) \; dt\right ] 
    \end{split}
\end{align}

Note that (\ref{causal-estimator-BNP-def-1}) does not adjust for potential confounding by censoring  given our assumption that censoring is noninformative conditional on the covariate and treatment assignment, and that probability of censoring is positive. Given the constructions described in (\ref{DSBP-construct-def0}--\ref{dependent-mix-eq-2}), we obtain 
\begin{align}
    \begin{split}
    \label{causal-estimator-BNP-def-2}
        \widehat{\Delta} 
        &= \frac{1}{N} \sum_{i=1}^N 
        E_{\bm{\theta}} \left[\int_0^\tau S_{\bm{\theta}}(t \; | \;A=1, \bm{X}_i=\bm{x}_i) - S_{\bm{\theta}}(t \; | \;A=0, \bm{X}_i=\bm{x}_i) \; dt\right ] 
         \\ &= \frac{1}{N} \sum_{i=1}^N E_{\bm{\theta}}\Bigg[\sum_{h=1}^L \Big[ \pi_h\big(\bm{\psi}_{A1,1}(\bm{w}_i)^\prime\bm{\alpha}_h\big) \int_0^\tau \int_t^\infty K\big(s \; | \; \bm{\psi}_{A1,2}(\bm{w}_i)^\prime\bm{\beta}_h,\bm{\omega}_h\big) \; ds dt \\ & - \pi_h\big(\bm{\psi}_{A0,1}(\bm{w}_i)^\prime\bm{\alpha}_h\big) \int_0^\tau \int_t^\infty K\big(s \; | \; \bm{\psi}_{A0,2}(\bm{w}_i)^\prime\bm{\beta}_h,\bm{\omega}_h\big) \; ds dt \Big]\Bigg]
    \end{split}
\end{align}
where $1\leq L \leq \infty$, $\bm{\theta}=\{\bm{\alpha},\bm{\beta},\bm{\omega}\}$, and $\bm{\psi}_{\cdot,\cdot}(\cdot)^\prime$ is a linear transformation function. Consequently, a $100(1-\alpha)\%$-level credible interval (CI) for $\Delta$ is given by $[\widehat{\Delta}^{(\alpha/2)},\widehat{\Delta}^{(1-\alpha/2)}]$ where $\widehat{\Delta}^{(\alpha/2)}$ and $\widehat{\Delta}^{(1-\alpha/2)}$ are calculated as the $(\alpha/2)$th and $(1-\alpha/2)$th quantiles of the posterior distribution of $\Delta$. The closed form equations of (\ref{subj-estimator-def-2}) assuming a Weibull and gamma kernel density, respectively, are shown in the appendix.

\subsection{Prior Specifications for Posterior Sampling of Proposed Approaches \label{chap3-posterior-scheme}}


Given sample observations $ \big\{o_i =\big(y_i,\bm{w}_i=(a_i,\bm{x}_i),\delta_i\big):i=1,\ldots,N\big\}$, we specify the likelihood and priors of the DSBP prior mixture models with covariates dependence on both the mixing probabilities and kernel densities in a hierarchical representation. For the $h^{th}$ cluster,
\begin{align}
    \begin{split}
    \label{gSB-mix-mod-spec-1}
    &y_i \ | \ \bm{\beta}_{i,h}, \omega_h \stackrel{i.i.d.}{\sim} K\big(y_i \; | \; exp\big\{\bm{\psi}(\bm{w}_i)^\prime\bm{\beta}_{i,h}\big\}, \omega_{h}\big), \quad i=1,\ldots,N \\
    &v_h(\bm{w}_i) = g(\bm{\psi}(\bm{w}_i)^\prime \alpha_h), \; \pi_1(\bm{w}_i)=v_1(\bm{w}_i) \\  &\pi_h(\bm{w}_i)=\big(1-v_1(\bm{w}_i)\big)\big(1-v_2(\bm{w}_i)\big)\ldots\big(1-v_{h-1}(\bm{w}_i)\big)v_h(\bm{w}_i), \quad h \in\{2,\ldots,L-1\} \\
        &\bm{\alpha}_h=\{\alpha_{h,A},\alpha_{h,1},\ldots,\alpha_{h,R}\} \sim \mathcal{N}_{R+1}(\bm{\mu}_\alpha,\bm{\Sigma}_\alpha), \quad h \in \{1,\ldots,L-1\} \\
        &\bm{\beta}_h=\{\beta_{h,A},\beta_{h,1},\ldots,\beta_{h,R}\} \sim \mathcal{N}_{R+1}(\bm{\mu}_\beta,\bm{\Sigma}_\beta), \quad h \in \{1,\ldots,L\} 
        \\ &\omega_h \sim unif(c_{low},c_{up}), \quad h \in \{1,\ldots,L\}
    \end{split}
\end{align}

where $(\bm{\mu}_\alpha,\bm{\Sigma}_\alpha,\bm{\mu}_\beta,\bm{\Sigma}_\beta,c_{low},c_{up})$ are constants, and $g(\cdot)$ can be a standard probit (regression) model, a shrinkage probit model (\ref{SPSBP-prior-def1}), or an inverse logit (regression) model. We can choose $K(y_i \; | \; \cdot)$ to be either a Weibull kernel density (scale$=exp\big\{\bm{\psi}(\bm{w}_i)^\prime\bm{\beta}_{ih}\big\}$, shape$=\omega_{h}$) or a gamma kernel density (rate$=exp\big\{\bm{\psi}(\bm{w}_i)^\prime\bm{\beta}_{ih}\big\}$, shape$=\omega_{h}$) whose support is on $\mathbb{R}^+$. We reparameterize $\bm{\psi}(\bm{w}_i)^\prime\bm{\beta}_{ih}$ on an exponential scale in order to select priors with support on $\mathbb{R}^{+}$. The $\bm{\alpha}$ matrix is of dimension $N \times (L-1)$ since $\pi(\bm{w})$ has $L-1$ degrees of freedom ($\pi_{L}(\bm{w})=1-\sum_{h=1}^{L-1} \pi_h(\bm{w})$). Let $\mathcal{Q}(\cdot)$ denote the log joint density function, and let $h(\cdot)$ and $H(\cdot)$ denote the density and cumulative distribution function of the censoring r.v., respectively. Assuming a non-informative right censoring mechanism,
\begin{align*}
\begin{split}
    &\mathcal{Q}\big(\bm{\theta}=(\bm{\alpha},\bm{\beta},\bm{\omega}),\bm{o}\big) = \prod_{i=1}^N \Big\{\big[f_{\bm{\theta}}(y_i)\big(1-H(y_i)\big)\big]^{\delta_i}\big[S_{\bm{\theta}}(y_i) h(y_i)\big]^{(1-\delta_i)}\Big\} 
    \\
    &\propto \prod_{i=1}^N \sum_{j=1}^{L} \Big\{v_j(\bm{w}_i) \prod_{l < j}\big[1-v_l(\bm{w}_i)\big]\Big\} \cdot \Big[ K\big(y_i \; | \; exp\big\{\bm{\psi}(\bm{w}_i)^\prime\bm{\beta}_{i,j}\big\}, \omega_{j}\big) \Big]^{\delta_i} \\ &\cdot \Big[ \int_t^\infty K\big(y_i \; | \; exp\big\{\bm{\psi}(\bm{w}_i)^\prime\bm{\beta}_{i,j}\big\}, \omega_{j}\big) \Big]^{(1-\delta_i)} \propto \prod_{i=1}^N \sum_{j=1}^{L} \Big\{g(\bm{\psi}(\bm{w}_i)^\prime \bm{\alpha}_{i,j}) \prod_{l < j}\big[1-g(\bm{\psi}(\bm{w}_i)^\prime \bm{\alpha}_{i,j})\big] \Big\} \\ &\cdot \Big[ K\big(y_i \; | \; exp\big\{\bm{\psi}(\bm{w}_i)^\prime\bm{\beta}_{i,j}\big\}, \omega_{j}\big) \Big]^{\delta_i} \cdot \Big[ \int_t^\infty K\big(y_i \; | \; exp\big\{\bm{\psi}(\bm{w}_i)^\prime\bm{\beta}_{i,j}\big\}, \omega_{j}\big) \Big]^{(1-\delta_i)}
\end{split}
\end{align*}

where $\bm{\alpha}=\{\bm{\alpha}_{i,j} \; | \; i=1,\ldots,N,j=1,\ldots,L-1\}; \; \bm{\alpha}_{i,j}=\{\bm{\alpha}_{i,j,1},\ldots,\bm{\alpha}_{i,j,R}\}$, $\bm{\beta}=\{\bm{\beta}_{i,j} \; | \; i=1,\ldots,N,j=1,\ldots,L\}; \; \bm{\beta}_{i,j}=\{\bm{\beta}_{i,j,1},\ldots,\bm{\beta}_{i,j,R}\}$, $\bm{\omega}=\{\omega_1,\ldots,\omega_L\}$, $\bm{o}=\{\bm{o}_1,\ldots,\bm{o}_N\}; \; o_i=\{y_i,\delta_i,\bm{w}_i\}$.




\section{Simulation Studies \label{BNP-gSB-chap-sim-study}}

\subsection{Survival and Censoring Time Data Generation Models}

We consider two data generation settings: ($i$) randomized controlled trial (RCT) with a balanced design; ($ii$) observational study where treatment assignments are confounded by observed covariates. We focus on simulating time-to-event data using various data generative models at different sample size levels, e.g., $N=\{200,500,1,000\}$. We consider a non-informative right-censoring mechanism with two components: ($i$) all patients are subject to random censoring/drop-out after enrollment; ($ii$) all patients who don't experience an event are censored after a maximum follow-up time. Hence, we assign a time-to-censoring random variable with an exponential distribution $C \sim exp(rate=\lambda_C=0.05)$ and censor all observations beyond a maximum follow-up time of $\tau_c=10.5$ years. We also incorporate a random recruitment mechanism ($2$-year period) following a uniform distribution of ($0,2$). We independently generate three covariates of mixed-type ($2$ continuous and $1$ discrete): $X_1 \sim \mathcal{N}(0,1)$, $X_2 \sim Binomial(N,p)$, and $X_3 \sim t(d.f.=5,ncp=0)$. Therefore, the observed data is of structure: $o_i=(a_i,x_{i,1},x_{i,2},x_{i,3},y_i,\delta_i), \ \text{for} \ i=1,\ldots,N; \; \bm{o}=\{o_1,\ldots,o_N\}$ where $y_i=min(c_i,t_i,\tau_c)$ and $\delta_i=\bone\{t_i \leq c_i\}$. For notational convenience, we denote the tuple of covariates and treatment assignment for the $i^{th}$ patient by $w_i=(a_i,x_i)$. For the RCT setting, we randomly assign patients, with equal probability of $50\%$ ($A \sim Binomial(N,p=0.5)$), to one of the two treatment groups: a test and a control group denoted by $A_1$ and $A_0$, respectively. For the observational study setting, we specify treatment assignment probabilities as a linear combination of covariates values under a logit transformation: $logit(p_i) = \bm{x}_i^T \bm{\beta}_i$ where $\bm{x}_i=\{x_{i,1},x_{i,2},x_{i,3}\}$ and $\bm{\beta}=\{\beta_1=1,\beta_2=1,\beta_3=1\}$ for $i=1,\ldots,N$ such that $A_i \sim Binom(N,p=\bm{x}_i^\prime \bm{\beta})$. Under the Weibull survival time generation model, the data generating model for survival time $T$ is a Weibull distribution with baseline shape parameter $\gamma_0$, baseline scale parameter $\lambda_0$ and multiplicative covariates effects. For the $i^{th}$ patient, its survival function is specified as:
\begin{equation}
    \label{RNG-weibull-1}
    S_i(t) = exp\{-\lambda_0 (t^{\gamma_0}) exp(\bm{w}_i^\prime \bm{\beta}_{Weibull})\}
\end{equation}


where $\bm{w}_i=\{a_i,x_{i,1},x_{i,2},x_{i,3}\}$, $\bm{\beta}_{Weibull}=\{\beta_a=-2.5,\beta_{x1}=1.5,\beta_{x2}=2.5,\beta_{x3}=1.5\}$, $\lambda_0=1.5$, and $\gamma_0=1.5$. The true RMST value for the $i^{th}$ patient is evaluated by integrating (\ref{RNG-weibull-1}) from $t=0$ to $\tau$ given the patient's covariates values and coefficients of covariates effect ($\bm{\beta}_{Weibull}$). Under the lognormal survival time generation model, the survival time $T$ is assigned a lognormal distribution with a fixed standard deviation $(\sigma=1)$ and mean parameter that is a linear combination of covariates $(\mu = \bm{w}^\prime \bm{\beta})$. For the $i$th subject,
\begin{equation}
\label{RNG-Lognorm-1}
    t_i  \stackrel{i.i.d.}\sim Lognormal \; \big(log(\mu_i)=\bm{w}_i^\prime \bm{\beta}_{lognormal}, \ log(\sigma_i) = 1\big)
\end{equation}

where $\bm{w}_i=\{a_i,x_{i,1},x_{i,2},x_{i,3}\}$, and $\bm{\beta}_{lognormal}$ controls covariates' effect on survival. We assign $\bm{\beta}^1_{lognormal}=\{\beta_a=2.5,\beta_{x1}=1.5,\beta_{x2}=2.5,\beta_{x3}=1.5\}$ for a positive treatment effect (default setting), $\beta_{x1}=-1.5$, and $\beta_{x1}=0$ for a negative and null treatment effect, respectively. 
The true RMST value for the $i^{th}$ patient is evaluated by $\int_0^\tau 1- \Phi\big((t_i - \mu_i)\big) dt_i$ given the patient's covariates values and coefficients of covariates effect ($\bm{\beta}_{lognormal}$). Additionally, we consider a two-component Weibull mixture survival time generation model, which allows for more flexible baseline hazard functions. The two-component mixture Weibull distributions are additive on the survival scale, with a mixing proportion parameter $\rho$, i.e. $S(t) = \rho S_1(t) + (1-\rho) S_2(t)$. The survival function for the $i^{th}$ patient is defined as
\vspace{-.5em}
\begin{align}
    \begin{split}
    \label{RNG-2Weib-1}
        S_i(t) = \big( \rho \cdot e^{-\lambda_0^1 (t^{\gamma_0^1})} + (1-\rho) \cdot e^{-\lambda_0^2 (t^{\gamma_0^2})} \big)^{exp\{\bm{w}_i^\prime \bm{\beta}_{2Weibull}\}}
    \end{split}
\end{align}

where $\rho$ denotes the mixture proportion; ($\lambda_0^1,\gamma_0^1$), ($\lambda_0^2,\gamma_0^2$) denote the baseline scale and shape parameters for the two mixture distributions; $\bm{w}_i=\{a_i,x_{i,1},\ldots,x_{i,R}\}$ denotes observed treatment assignment and covariates for the $i^{th}$ subject; $\bm{\beta}_{2Weibull}$ denotes the coefficients of covariates effect. The true RMST value for the $i^{th}$ patient is evaluated by integrating \ref{RNG-2Weib-1} from $t=0$ to $\tau$ given $(\bm{w}_i,\bm{\beta}_{2Weibull})$ where the coefficients $\bm{\beta}_{2Weibull}$ are set at the same numerical values as $\bm{\beta}_{Weibull}$ and $\bm{\beta}^2_{lognormal}$.

\subsection{Simulation Scenarios}

For a comprehensive evaluation of our inferential tools, we consider a total of $6$ data generation scenarios: \{$3$ data generation models: Weibull, lognormal, and two-components Weibull mixture\}$\times$\{$2$ study settings: randomized controlled trial and observational study\}. The lognormal model has two extra covariates settings: negative and null treatment effects in addition to the default positive treatment effect setting shared by the Weibull and two-component Weibull mixture models. 
We evaluate performance of the proposed BNP estimators of RMST, and compare with two frequentist methods: ($i$) \cite{tian2014predicting}'s direct (RMST) regression method; ($ii$) \cite{ambrogi2022analyzing}'s pseudo-values method. We make evaluations in terms of bias, root mean square error (RMSE), and coverage probability (CP). 
We also evaluate Bayesian models' performance under different choices of DSBP priors and kernel densities. For each simulation scenario, we randomly generate and fix a sample of covariates ($\bm{W}$) at a given sample size. Then for each simulation replication, we randomly generate a sample of outcomes ($\bm{Y},\bm{\delta}$) given $\bm{W}$ and make inferences given observed data $\bm{O}=(\bm{W},\bm{Y},\bm{\delta})$. In an oncology study setting, $5$ year is usually considered a mile-stone for treatment evaluation. Hence for fixed-time analysis, our simulation studies evaluate RMST estimations at $\tau=5$ years. Besides, we conduct RMST curve estimations on a grid of time points. For prior specifications, we set $\big(\bm{\mu}_\alpha=\bm{0},\bm{\Sigma}_\alpha=400 \cdot  I_{4},\bm{\mu}_\beta=\bm{0},\bm{\Sigma}_\beta=400 \cdot I_{4},c_{low}=\{0.01,0.1\},c_{up}=\{10,15\}\big)$ (\ref{gSB-mix-mod-spec-1}) such that $\bm{\alpha}$ and $\bm{\beta}$ each follows a four-dimensional Gaussian distribution with an independent covariance structure and equal standard deviation of $20$ for each predictor. We use a minimum burn-in iteration size of $2,000$ and a minimum posterior sampling iteration size of $1,000$ for all NUTS runs. Initial values are provided by Stan as a default setting.

\subsection{Simulation Results}

\begin{figure}[H]
\centering
 \renewcommand{\baselinestretch}{1}
\caption[Single time point analysis scenario 1]{Single time point analysis scenario 1: ATE point estimates with 95\% credible intervals comparing SPSBP prior with PSBP prior (lognormal data generation model under a RCT setting; $\tau=5$)}
     \begin{subfigure}[b]{0.45\textwidth}
     \includegraphics[width=\textwidth]{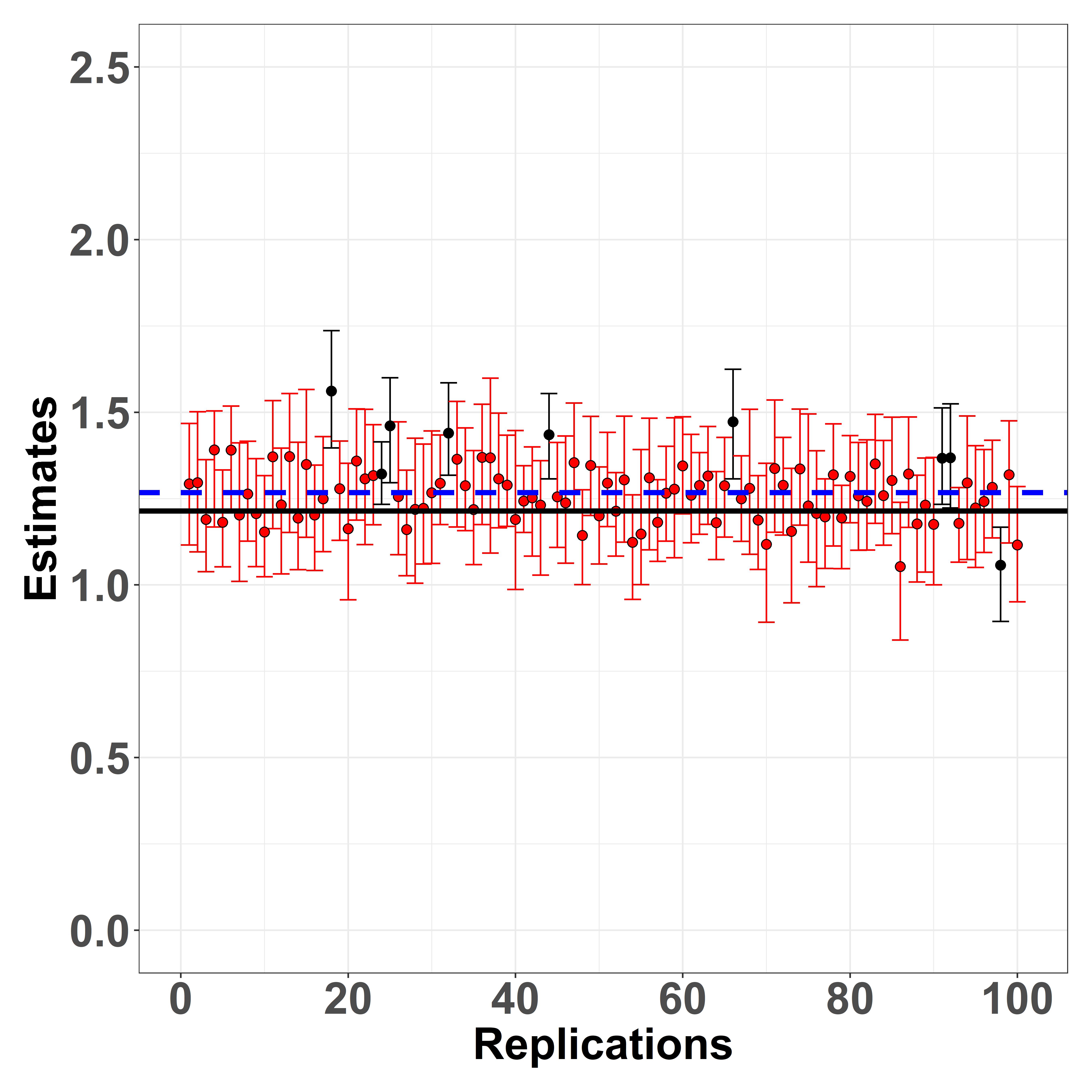}
     \caption{\centering SPSBP prior-Weibull kernel: CP = 0.91, Average CI Length=0.33, bias = 0.05, RMSE = 0.1}
     \end{subfigure}
     \begin{subfigure}[b]{0.45\textwidth}
     \includegraphics[width=\textwidth]{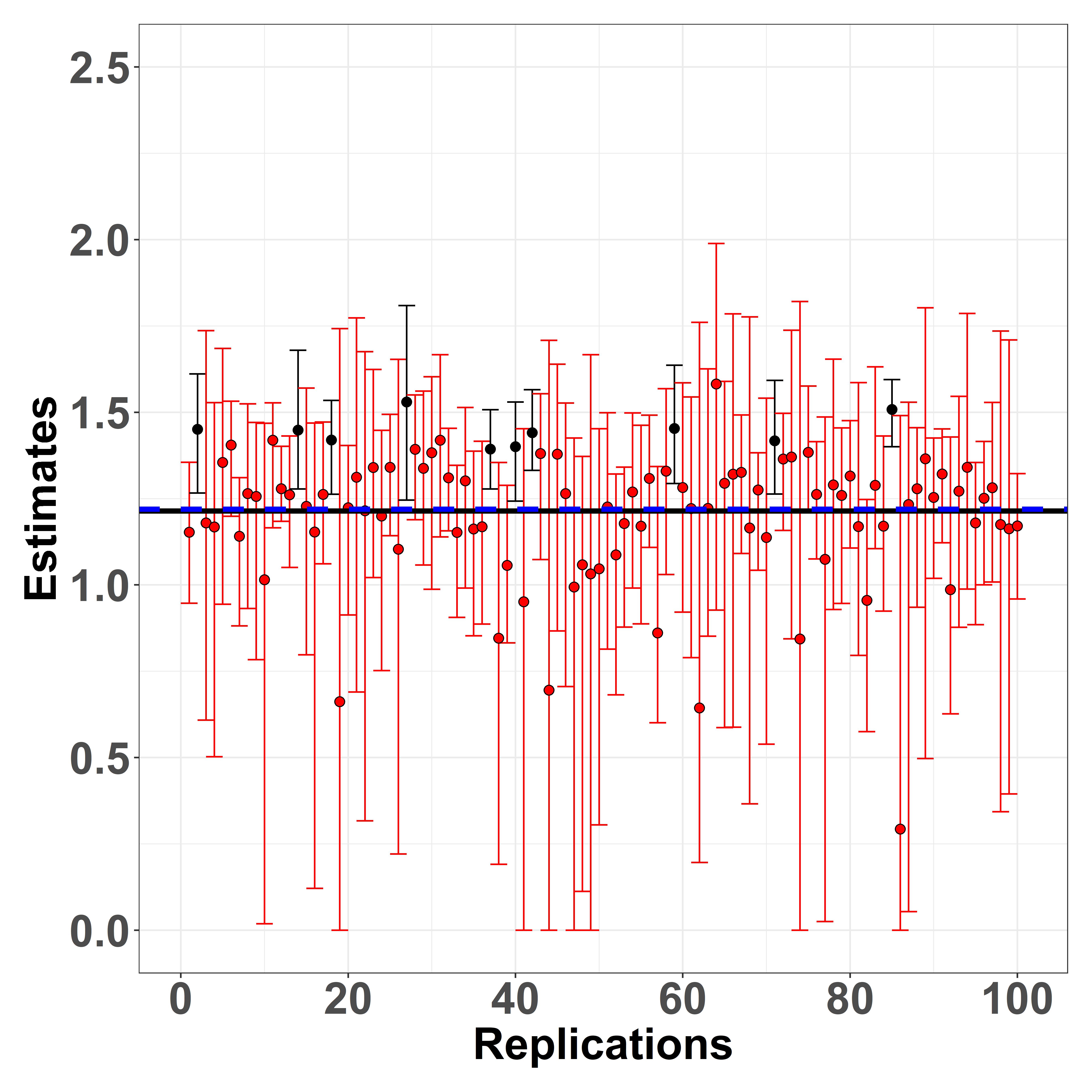}
     \caption{\centering PSBP prior-Weibull kernel: CP = 0.9, Average CI Length=0.74, bias = 0.01, RMSE = 0.2}
     \end{subfigure}
    \caption*{\small CP denotes coverage probability; ACredIntL denotes average credible intervals' length; sample size $N=500$; RMSTD evaluated at $\tau=5$ years; coefficients of predictor effects: $(\beta_{A}=2.5,\beta_{W_1}=1.5,\beta_{W_2}=2.5,\beta_{W_3}=1.3)$; black solid lines and blue dashed lines mark the true RMSTD value and average RMSTD point estimates, respectively.}
    \label{fig-chapter-3-fixed-tau-1}
\end{figure}

Figure \ref{fig-chapter-3-fixed-tau-1} shows estimation results on ATE by BNPDM models assuming a SPSBP prior and a PSBP prior, both with a Weibull kernel density, where survival data are generated by a lognormal model under a RCT setting. Dots and bars denote point estimates and credible intervals, respectively. A bar and a dot are colored red together if the credible interval covers the true/population ATE. Given a moderate sample size ($500$), biases (averaged over $100$ replications) of both models are near zero. However, credible intervals and RMSE estimated under the SPSBP prior are much tighter and smaller compared to those of the PSBP prior. Specifically, the average credible interval (CI) length of the SPSBP prior is less than half of that of the PSBP prior ($0.33/0.74 \approx 0.45$). The (default) PSBP prior model give volatile estimates while the modified PSBP prior (SPSBP prior) model stabilize estimates and results in ``shrinked'' credible intervals in comparison.

\begin{figure}[H]
\centering
 \renewcommand{\baselinestretch}{1}
\caption[Single time point analysis scenario 2]{Single time point analysis scenario 2: ATE point estimates with 95\% credible intervals comparing LSBP prior with LDDP prior (lognormal data generation model under a RCT setting; $\tau=5$)}
    \begin{subfigure}[b]{0.45\textwidth}
     \includegraphics[width=\textwidth]{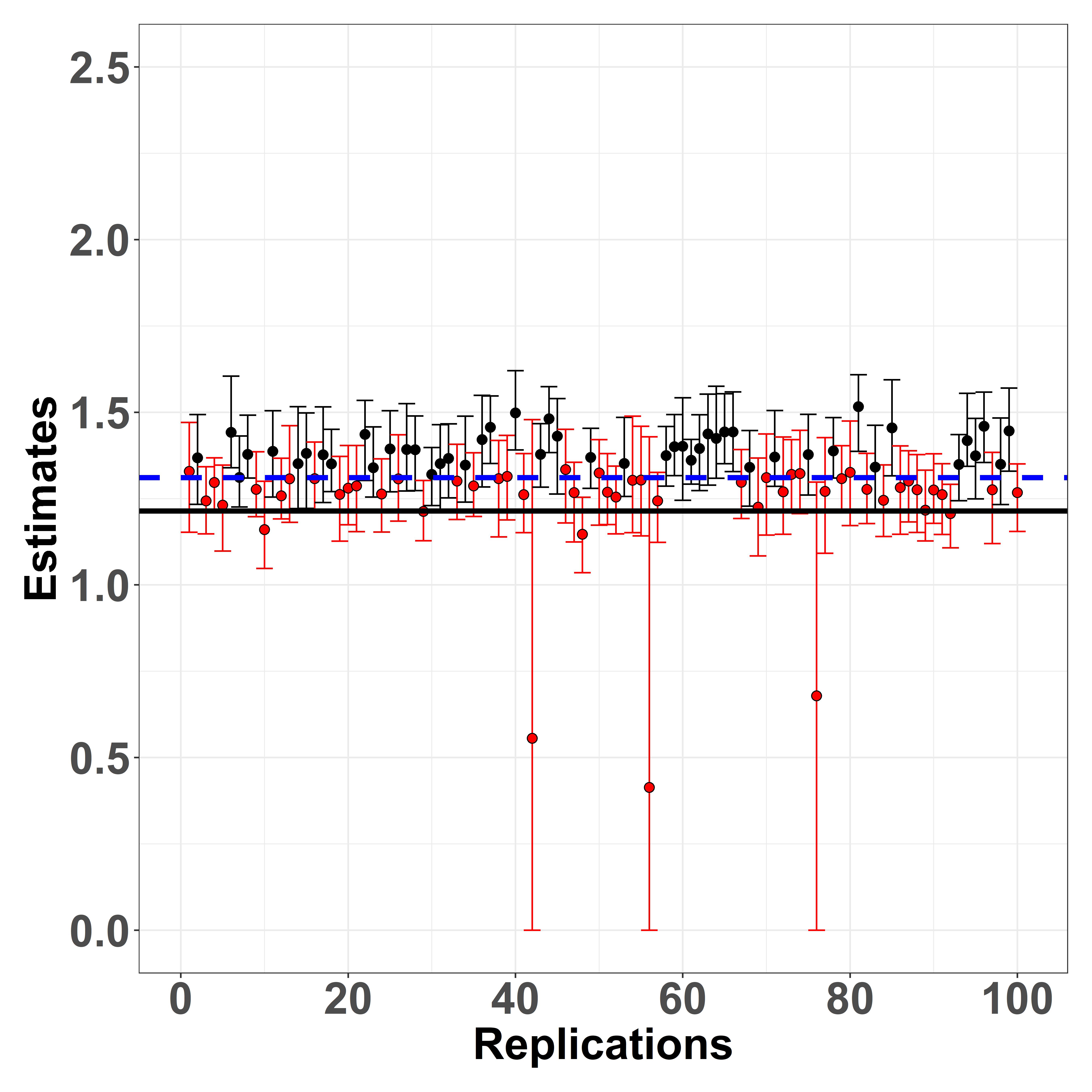}
     \caption{\centering LSBP prior-Weibull kernel: CP = 0.52, Average CI Length=0.26, bias=0.1, RMSE=0.18}
     \end{subfigure}
    \begin{subfigure}[b]{0.45\textwidth}
     \includegraphics[width=\textwidth]{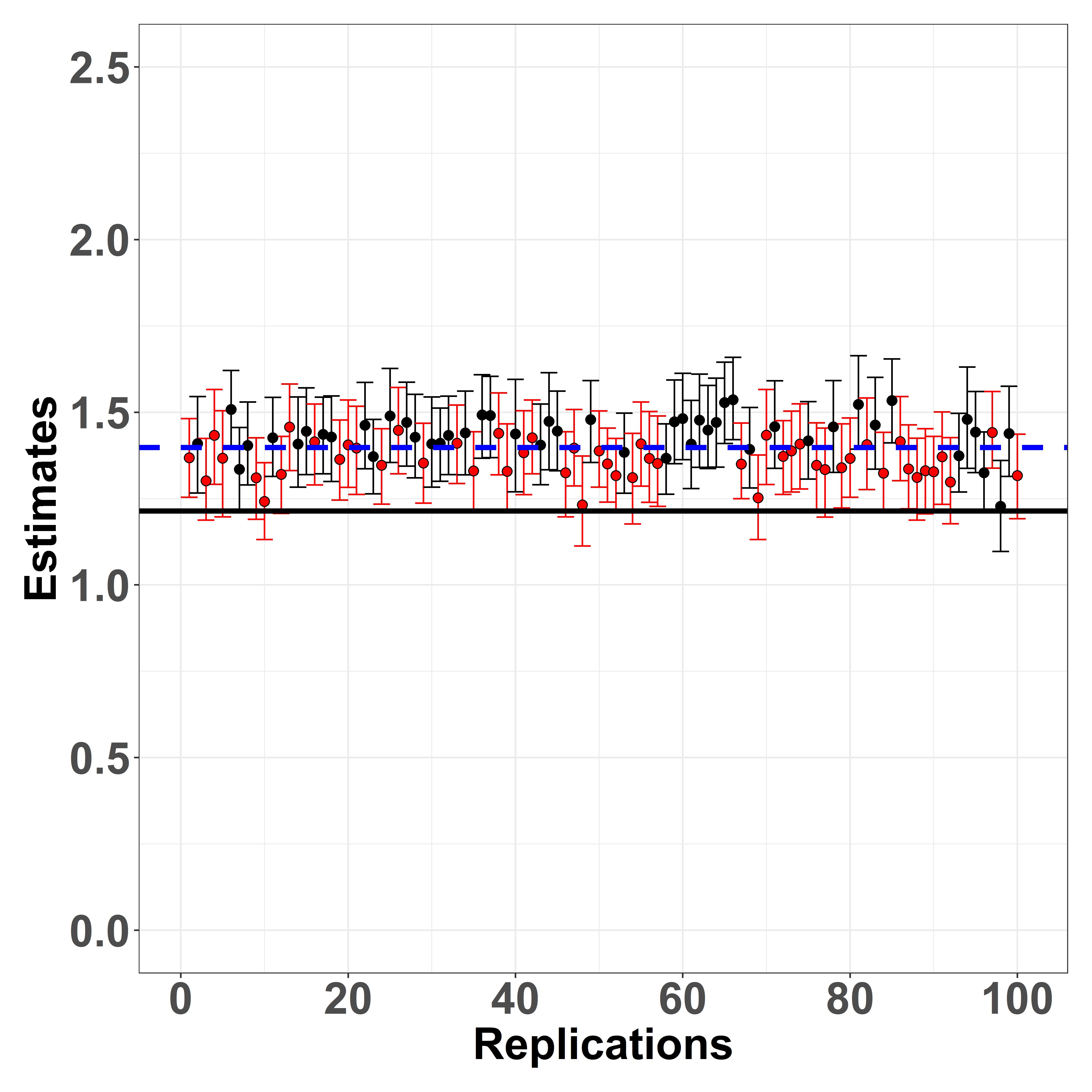}
     \caption{\centering LDDP prior-Weibull kernel: CP = 0.21, Average CI Length=0.24, bias=0.18, RMSE=0.2}
     \end{subfigure}
    \caption*{CP denotes coverage probability; ACredIntL denotes average credible intervals' length; sample size $N=500$; RMSTD evaluated at $\tau=5$ years; coefficients of predictor effects: $(\beta_{A}=2.5,\beta_{W_1}=1.5,\beta_{W_2}=2.5,\beta_{W_3}=1.3)$; black solid lines and blue dashed lines mark the true RMSTD value and average RMSTD point estimates, respectively.}
    \label{fig-chapter-3-fixed-tau-1b}
\end{figure}

Figure \ref{fig-chapter-3-fixed-tau-1b} shows results estimated by the LSBP prior and the LDDP prior models, both with a Weibull kernel density, under the same lognormal-RCT data generation setting. The LSBP and LDDP prior models result in higher biases and RMSEs, yet slightly smaller average CI length compared to the SPSBP prior (model). However, their CPs are very low ($52\%$ and $21\%$) compared to SPSBP prior and PSBP prior ($91\%$ and $90\%$) though the PSBP prior's CP may be inflated due to its extra wide credible intervals.

Regarding subject-level RMST inference, all DSBP prior (SPSBP, PSBP, and LSBP) models have superior performance compared to their frequentist counterparts \citep{tian2014predicting,ambrogi2022analyzing} under the two-components Weibull mixture data generation model as shown in Figure F1--F5 (supplemental materials). Numerical results (for $\tau=5$; Figure F1--F2) in Table \ref{tab-chap3-subjs-plot-1} show that, with a sample size of $500$, RMSE given by the SPSBP prior is less than half of that by \cite{tian2014predicting}'s method ($0.25$ compared to $0.57$). For subject-level CP, which is defined by the proportion of time the credible or confidence interval covers the true individual RMSTD value, the LSBP prior yields $0.88$. In comparison, CP by \cite{tian2014predicting} and \cite{ambrogi2022analyzing} are both $0.48$. Furthermore, this higher CP is not achieved with increased interval width. On the contrary, the average credible interval length of the LSBP prior is $0.42$ compared to $0.75$ (average confidence interval length) given by the frequentists' methods.

\begin{table}[H]
\centering
\caption[SUBJECT-LEVEL RMST PREDICTIONS SCENARIO 1-A]{
SUBJECT-LEVEL RMST PREDICTIONS SCENARIO 1-A}
      \renewcommand{\baselinestretch}{1}
    \setlength{\tabcolsep}{6pt} 
\renewcommand{\arraystretch}{1}
\vspace{5pt}
\resizebox{.7\columnwidth}{!}{%
\begin{tabular}{|c|c|c|c|c|}
\hline
\textbf{Method} &
  \textbf{Bias} &
  \textbf{RMSE} &
  \textbf{\begin{tabular}[c]{@{}c@{}}Coverage\\ Probability\end{tabular}} &
\textbf{\begin{tabular}[c]{@{}c@{}}Average Credible/\\ Confidence\\ Interval Length\end{tabular}} \\ \hline
SPSBP-Weibull & 0.08 & 0.25 & 0.556 & 0.3121 \\ \hline
PSBP-Weibull  & 0.01 & 0.19 & 0.886 & 0.3830 \\ \hline
LSBP-Weibull  & 0.04 & 0.21 & 0.880 & 0.4164 \\ \hline
LDDP-Weibull  & 0.18 & 0.37 & 0.402 & 0.2239 \\ \hline
\cite{tian2014predicting}          & 0.00 & 0.57 & 0.484 & 0.7477 \\ \hline
\cite{ambrogi2022analyzing}      & 0.00 & 0.57 & 0.484 & 0.7448 \\ \hline
\end{tabular}}
\vspace{5pt}
          \label{tab-chap3-subjs-plot-1}
\end{table}

Restricted mean survival time is a function of restricted time, and estimating RMST at only a single time point $\tau$ does not tell the whole story of temporal survival relationship. Therefore, we expand the time horizon to evaluate the performance of BNPDM models for RMST inference on a grid of time points ($\tau$s). We calculate point estimates and point-wise credible intervals
. On the other hand, given $\tau$ and posterior sample $\bm{\theta}$ drawn from $Pr(\bm{\theta} \; | \; \bm{Y}, \bm{\delta}, \bm{W})$, the RMST function is just a deterministic function of these quantities. Therefore, we can attain an entire RMST curve estimate with a single NUTS run. We estimated RMST curves under $100$ data replications and show their results in Figures F11--F20 (supplemental materials). The corresponding numerical results are summarized in Table \ref{RMSTD-curve-subj-level-res-table}.

\begin{table}[H]
\centering
\caption{Subject-level RMSTD Curve Inference Results; Data generation model is a 2-components Weibull mixture model}
\resizebox{.975\columnwidth}{!}{%
\begin{tabular}{|c|ccccc|ccccc|}
\hline
\multirow{3}{*}{$\bm{\tau}$} &
  \multicolumn{5}{c|}{\textbf{Average Absolute Bias}} &
  \multicolumn{5}{c|}{\textbf{RMSE}} \\ \cline{2-11} 
 &
  \multicolumn{1}{c|}{\textbf{\begin{tabular}[c]{@{}c@{}}SPSBP\\ Weibull\end{tabular}}} &
  \multicolumn{1}{c|}{\textbf{\begin{tabular}[c]{@{}c@{}}LSBP\\ Weibull\end{tabular}}} &
  \multicolumn{1}{c|}{\textbf{\begin{tabular}[c]{@{}c@{}}LSBP\\ Gamma\end{tabular}}} &
  \multicolumn{1}{c|}{\textbf{\begin{tabular}[c]{@{}c@{}}Tian et al.\\ (2014)\end{tabular}}} &
  \textbf{\begin{tabular}[c]{@{}c@{}}Ambrogi et al.\\ (2022)\end{tabular}} &
  \multicolumn{1}{c|}{\textbf{\begin{tabular}[c]{@{}c@{}}SPSBP\\ Weibull\end{tabular}}} &
  \multicolumn{1}{c|}{\textbf{\begin{tabular}[c]{@{}c@{}}LSBP\\ Weibull\end{tabular}}} &
  \multicolumn{1}{c|}{\textbf{\begin{tabular}[c]{@{}c@{}}LSBP\\ Gamma\end{tabular}}} &
  \multicolumn{1}{c|}{\textbf{\begin{tabular}[c]{@{}c@{}}Tian et al.\\ (2014)\end{tabular}}} &
  \textbf{\begin{tabular}[c]{@{}c@{}}Ambrogi et al.\\ (2022)\end{tabular}} \\ \hline
\textbf{1} &
  \multicolumn{1}{c|}{0.67} &
  \multicolumn{1}{c|}{0.04} &
  \multicolumn{1}{c|}{0.10} &
  \multicolumn{1}{c|}{0.08} &
  0.08 &
  \multicolumn{1}{c|}{1.2} &
  \multicolumn{1}{c|}{0.06} &
  \multicolumn{1}{c|}{0.13} &
  \multicolumn{1}{c|}{0.10} &
  0.10 \\ \hline
\textbf{2} &
  \multicolumn{1}{c|}{0.53} &
  \multicolumn{1}{c|}{0.06} &
  \multicolumn{1}{c|}{0.11} &
  \multicolumn{1}{c|}{0.17} &
  0.17 &
  \multicolumn{1}{c|}{1.04} &
  \multicolumn{1}{c|}{0.10} &
  \multicolumn{1}{c|}{0.17} &
  \multicolumn{1}{c|}{0.21} &
  0.21 \\ \hline
\textbf{3} &
  \multicolumn{1}{c|}{0.42} &
  \multicolumn{1}{c|}{0.08} &
  \multicolumn{1}{c|}{0.14} &
  \multicolumn{1}{c|}{0.28} &
  0.28 &
  \multicolumn{1}{c|}{0.86} &
  \multicolumn{1}{c|}{0.14} &
  \multicolumn{1}{c|}{0.24} &
  \multicolumn{1}{c|}{0.34} &
  0.34 \\ \hline
\textbf{4} &
  \multicolumn{1}{c|}{0.34} &
  \multicolumn{1}{c|}{0.10} &
  \multicolumn{1}{c|}{0.16} &
  \multicolumn{1}{c|}{0.37} &
  0.37 &
  \multicolumn{1}{c|}{0.70} &
  \multicolumn{1}{c|}{0.18} &
  \multicolumn{1}{c|}{0.30} &
  \multicolumn{1}{c|}{0.47} &
  0.47 \\ \hline
\textbf{5} &
  \multicolumn{1}{c|}{0.28} &
  \multicolumn{1}{c|}{0.12} &
  \multicolumn{1}{c|}{0.18} &
  \multicolumn{1}{c|}{0.44} &
  0.44 &
  \multicolumn{1}{c|}{0.55} &
  \multicolumn{1}{c|}{0.23} &
  \multicolumn{1}{c|}{0.36} &
  \multicolumn{1}{c|}{0.57} &
  0.57 \\ \hline
\textbf{6} &
  \multicolumn{1}{c|}{0.24} &
  \multicolumn{1}{c|}{0.14} &
  \multicolumn{1}{c|}{0.20} &
  \multicolumn{1}{c|}{0.50} &
  0.50 &
  \multicolumn{1}{c|}{0.42} &
  \multicolumn{1}{c|}{0.27} &
  \multicolumn{1}{c|}{0.41} &
  \multicolumn{1}{c|}{0.67} &
  0.67 \\ \hline
\textbf{7} &
  \multicolumn{1}{c|}{0.21} &
  \multicolumn{1}{c|}{0.15} &
  \multicolumn{1}{c|}{0.21} &
  \multicolumn{1}{c|}{0.55} &
  0.55 &
  \multicolumn{1}{c|}{0.33} &
  \multicolumn{1}{c|}{0.31} &
  \multicolumn{1}{c|}{0.46} &
  \multicolumn{1}{c|}{0.75} &
  0.75 \\ \hline
\textbf{8} &
  \multicolumn{1}{c|}{0.19} &
  \multicolumn{1}{c|}{0.16} &
  \multicolumn{1}{c|}{0.23} &
  \multicolumn{1}{c|}{0.58} &
  0.59 &
  \multicolumn{1}{c|}{0.29} &
  \multicolumn{1}{c|}{0.36} &
  \multicolumn{1}{c|}{0.51} &
  \multicolumn{1}{c|}{0.83} &
  0.83 \\ \hline
\textbf{9} &
  \multicolumn{1}{c|}{0.21} &
  \multicolumn{1}{c|}{0.18} &
  \multicolumn{1}{c|}{0.24} &
  \multicolumn{1}{c|}{0.61} &
  0.62 &
  \multicolumn{1}{c|}{0.32} &
  \multicolumn{1}{c|}{0.40} &
  \multicolumn{1}{c|}{0.55} &
  \multicolumn{1}{c|}{0.90} &
  0.90 \\ \hline
\textbf{10} &
  \multicolumn{1}{c|}{0.23} &
  \multicolumn{1}{c|}{0.19} &
  \multicolumn{1}{c|}{0.25} &
  \multicolumn{1}{c|}{0.64} &
  0.65 &
  \multicolumn{1}{c|}{0.39} &
  \multicolumn{1}{c|}{0.44} &
  \multicolumn{1}{c|}{0.59} &
  \multicolumn{1}{c|}{0.96} &
  0.96 \\ \hline
\end{tabular}
}
\label{RMSTD-curve-subj-level-res-table}
\end{table}

\section{Real Data Applications \label{BNP-gSB-chap-real-data}}

The epidermal growth factor receptor (EGFR) has been proven to be a clinically meaningful target for monoclonal antibodies (mAbs) with efficacy established in treatment of metastatic colorectal cancer (mCRC) \citep{cunningham2004cetuximab,bokemeyer2009fluorouracil}. Panitumumab (Pmab) is a (fully) human anti-EGFR that was approved as monotherapy for patients with chemotherapy-refractory mCRC \citep{giusti2007fda}. A randomized phase III study was designed and conducted to evaluate the efficacy and safety of Pmab plus infusional fluorouracil, leucovorin, and oxaliplatin (FOLFOX4) versus FOLFOX4 alone as an initial treatment for mCRC in patients with previously untreated mCRC according to tumor \textit{KRAS} status \citep{douillard2010randomized}. The presence of activating \textit{KRAS} mutations was identified as a potent predictor of resistance to EGFR-directed antibodies (e.g., cetuximab and panitumumab) \citep{heinemann2009clinical}. In this real data application example we only focus on RMSTD inference among \textit{KRAS} wild type (WT) patients for its clinical relevance.

\begin{figure}[H]
\centering
 \renewcommand{\baselinestretch}{1}
 \centering
\caption[RMSTD curve estimation: real data analysis]{RMSTD curve estimation: real data analysis--comparing average treatment effect difference between FOLFOX4+Panitumumab versus FOLFOX 4 using PRIME \citep{douillard2010randomized} data (unit: months, sample size: $N=1,092$)}
 \begin{subfigure}[b]{0.3\textwidth}
    \caption{\small LSBP-Weibull}
     \includegraphics[width=\textwidth]{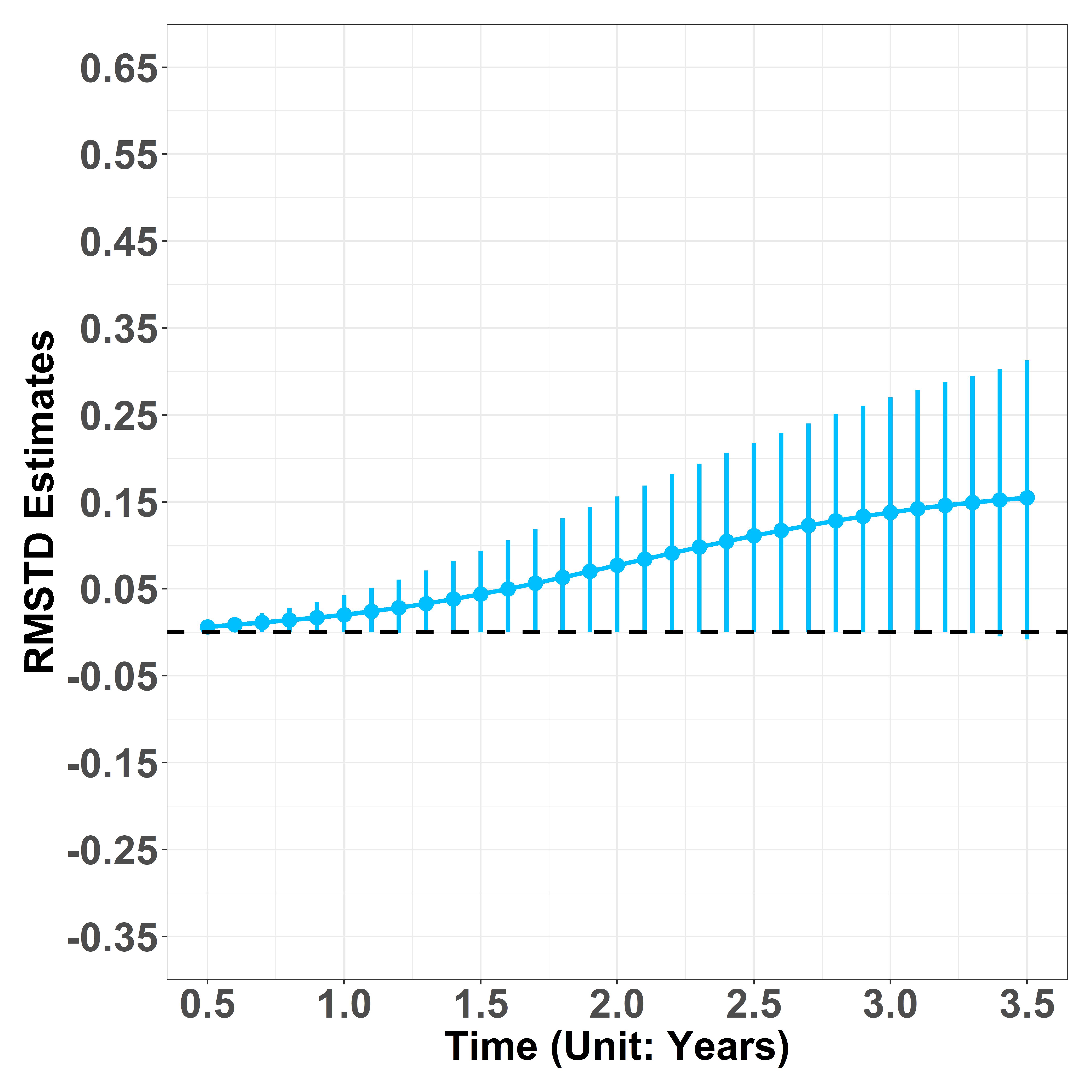}
     \end{subfigure}
    \begin{subfigure}[b]{0.3\textwidth}
    \caption{\small SPSBP-Weibull}
     \includegraphics[width=\textwidth]{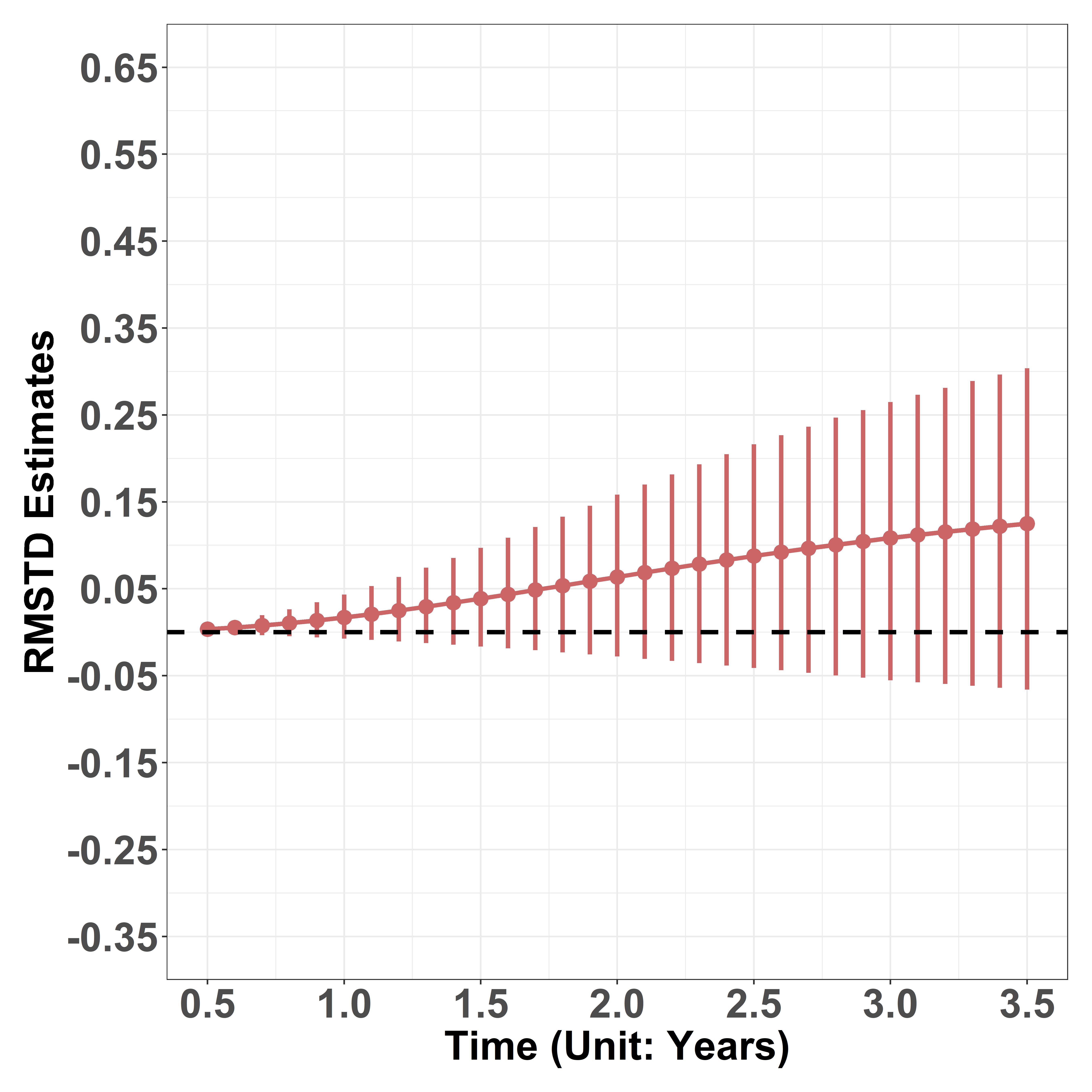}
     \end{subfigure}
         \begin{subfigure}[b]{0.3\textwidth}
    \caption{\small PSBP-Weibull}
     \includegraphics[width=\textwidth]{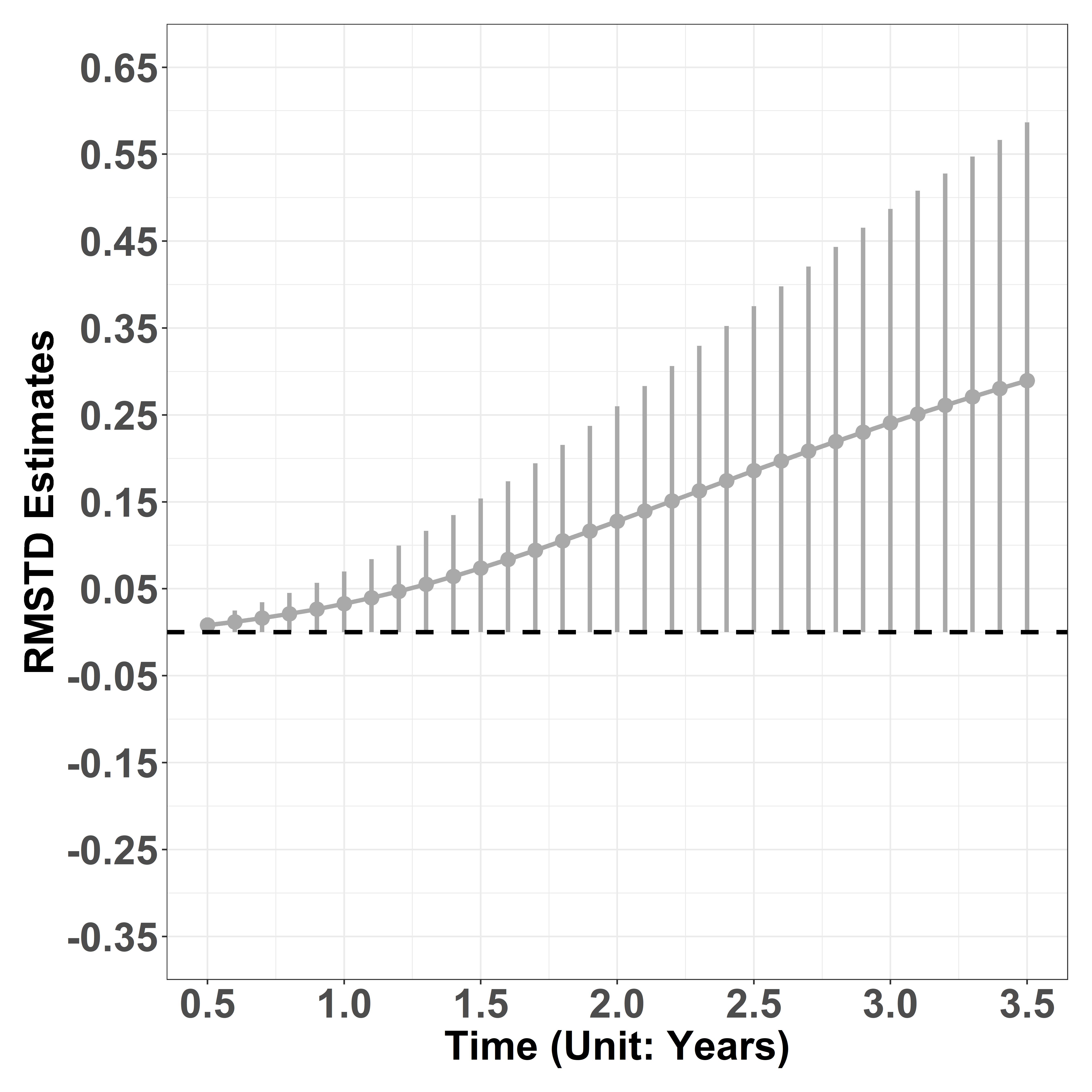}
     \end{subfigure}
         \begin{subfigure}[b]{0.3\textwidth}
    \caption{\small LSBP-Gamma}
     \includegraphics[width=\textwidth]{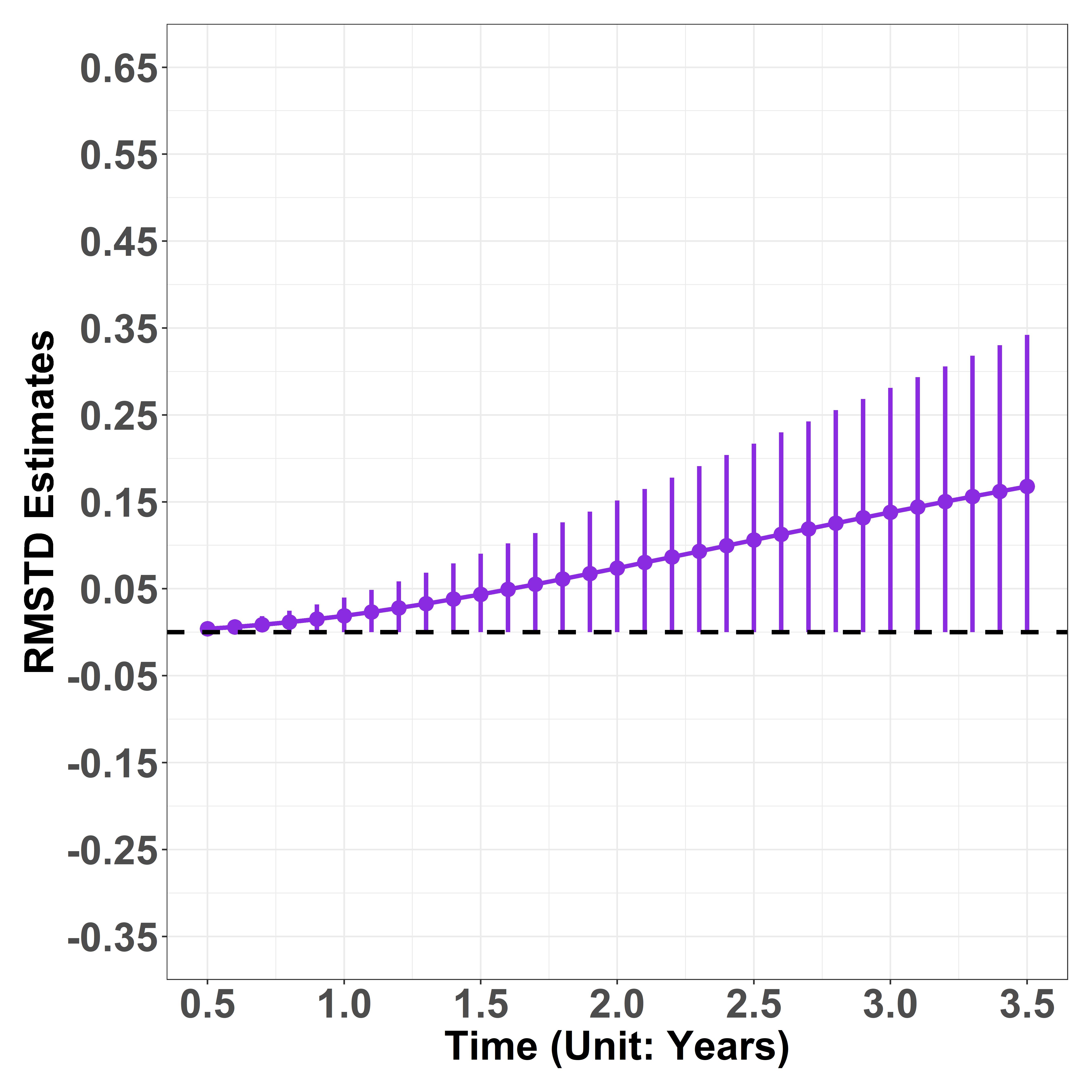}
     \end{subfigure}
    \begin{subfigure}[b]{0.3\textwidth}
    \caption{\small SPSBP-Gamma}
     \includegraphics[width=\textwidth]{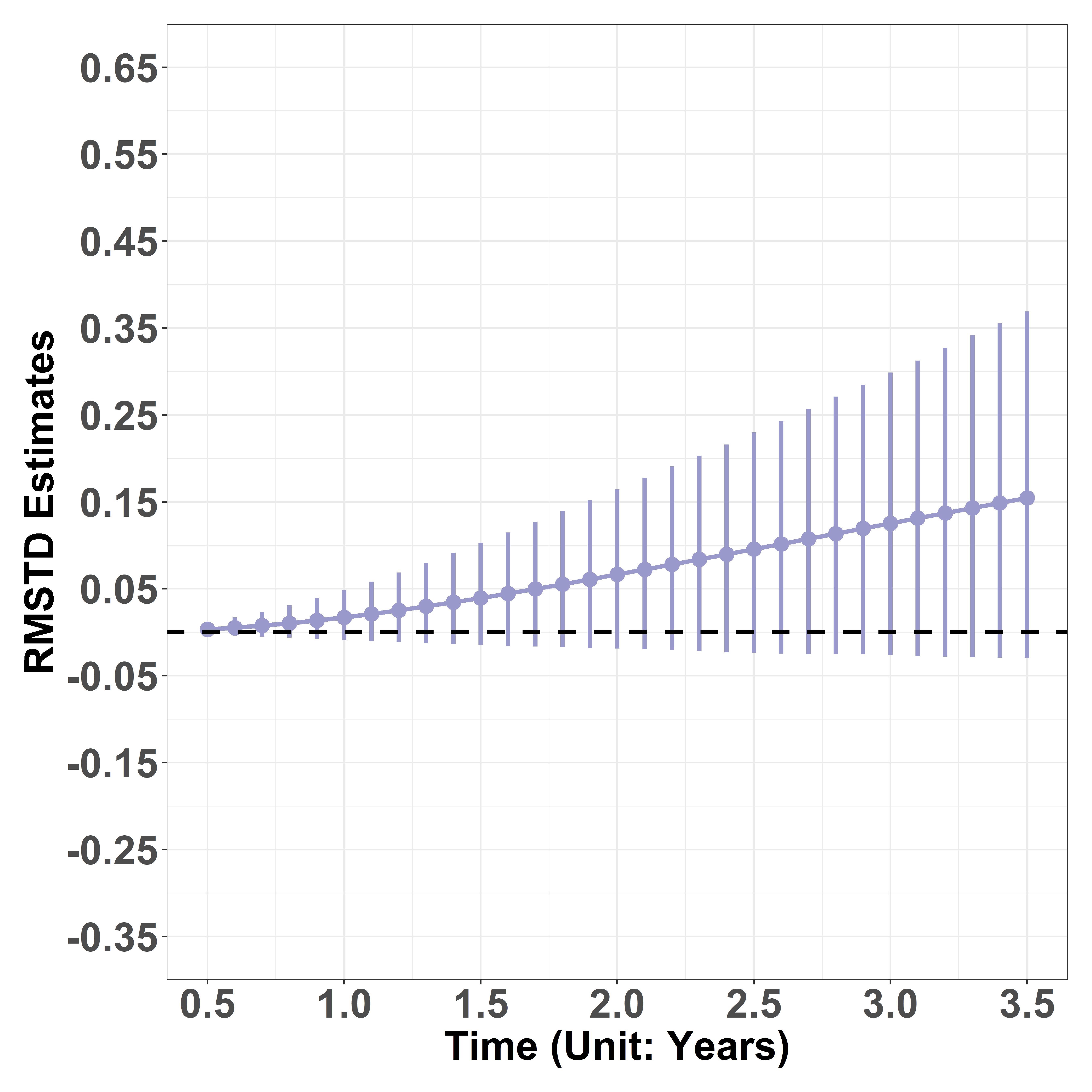}
     \end{subfigure}
    \begin{subfigure}[b]{0.3\textwidth}
    \caption{\small PSBP-Gamma}
     \includegraphics[width=\textwidth]{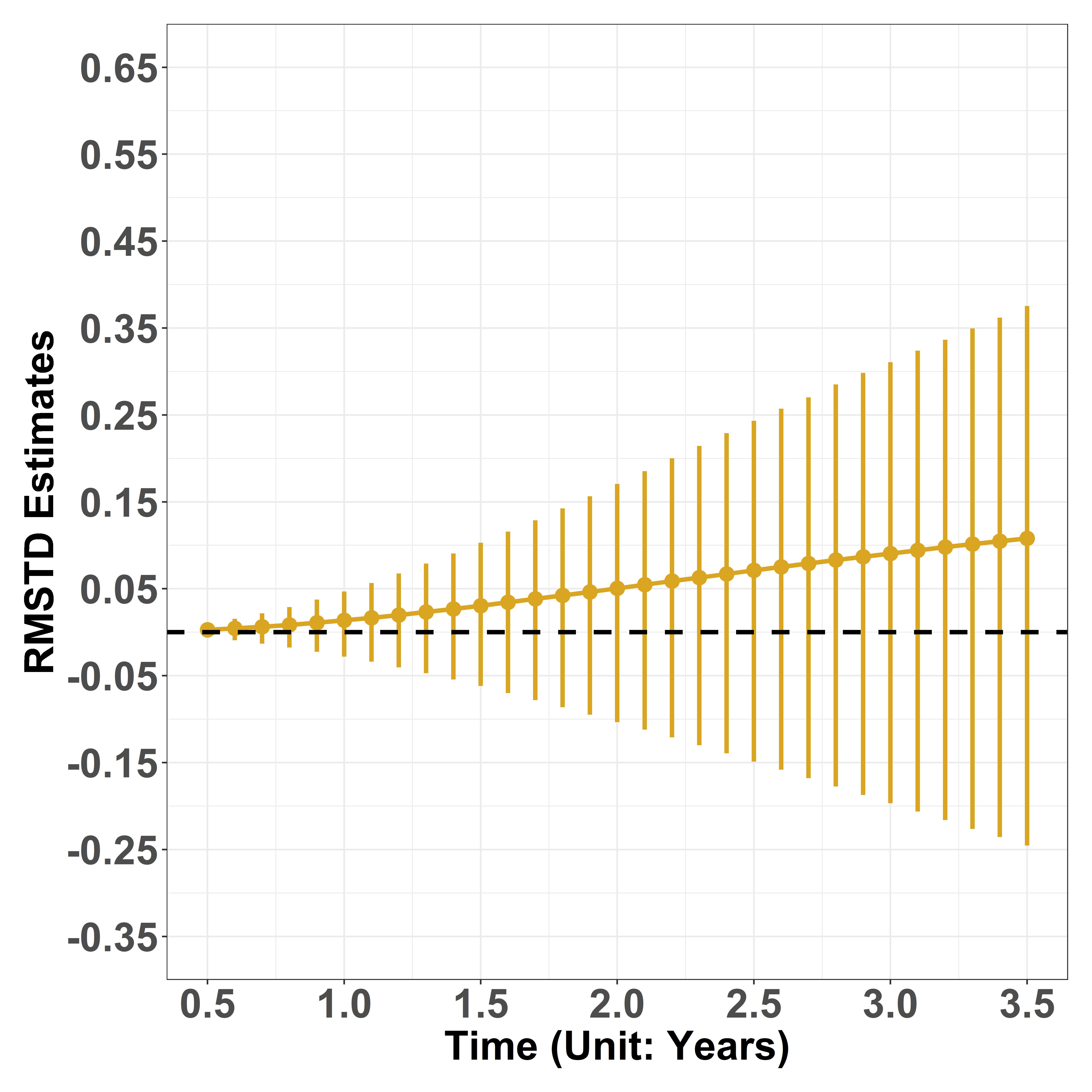}
     \end{subfigure}
         \begin{subfigure}[b]{0.3\textwidth}
    \caption{\small \cite{tian2014predicting}}
     \includegraphics[width=\textwidth]{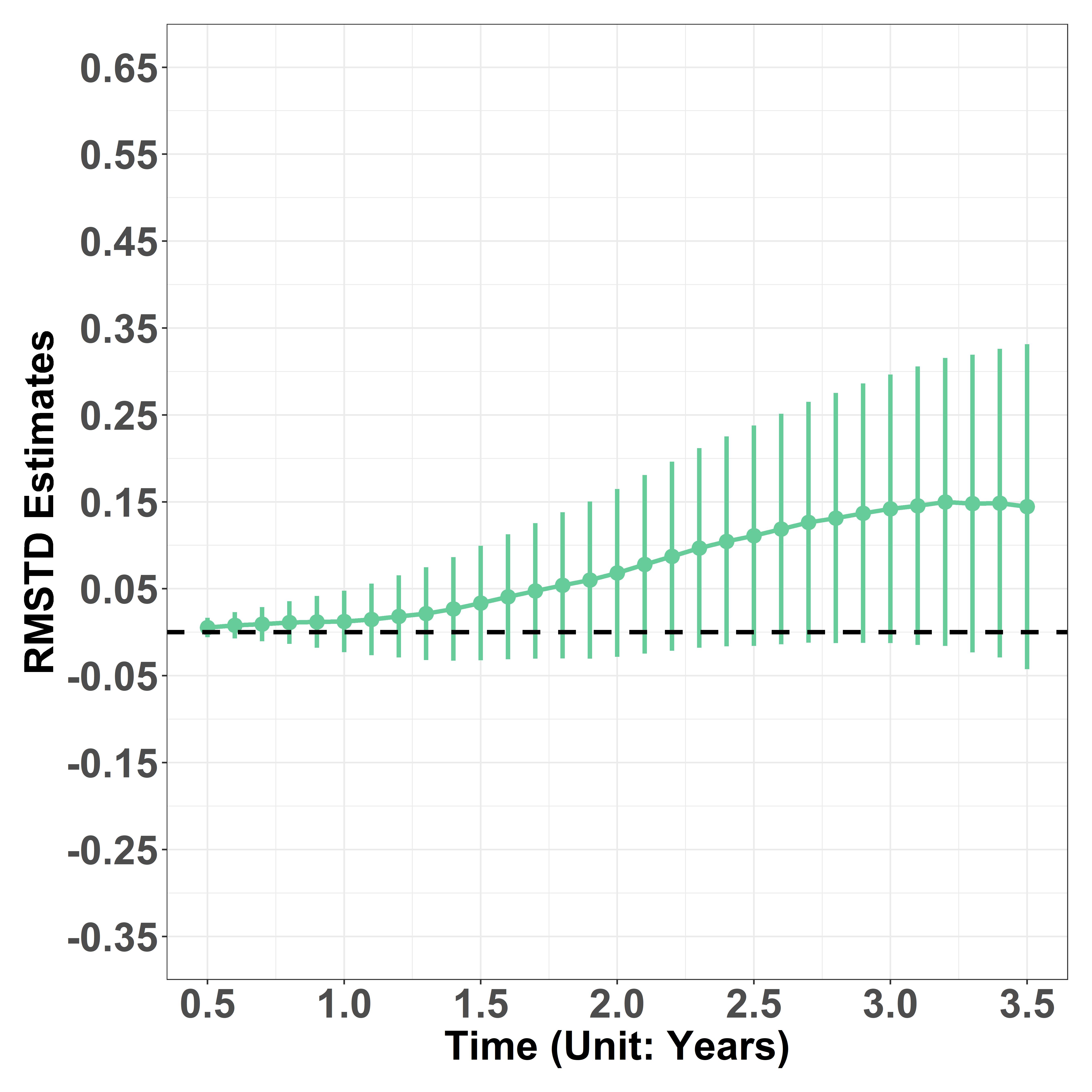}
     \end{subfigure}
    \label{fig-chapter-3-real-data-2}
\end{figure}

This phase III study was designed as an open-label, randomized, phase III trial to compare the treatment effect of adding Pamb to FOLFOX4 in patients with WT \textit{KRAS} tumors and also in patients with mutant (MT) \textit{KRAS} tumors \citep{douillard2010randomized}. A primary analysis of log-rank tests, stratified by random assignment factors, were conducted on the progression-free survival (PFS) and overall survival (OS) endpoints among the WT \textit{KRAS} and MT \textit{KRAS} patient stratums \citep{douillard2010randomized}. A prespecified final analysis, which included OS, was later reported in \citep{douillard2014final}. We conduct a reanalysis of the selected study to estimate group differential treatment effect on the OS endpoint. The maximum observed event time is approximately $3.8$ years, and we evaluate RMSTD up to $3.5$ years. We include BMI and age as predictive covariates, which are both prognostic in mCRC \citep{lieu2014association}. With all incomplete records removed, the study population has a total number of $652$ observations. We apply BNPDM models with SPSBP and LSBP priors both assuming a Weibull and Gamma kernel densities. We obtained both point estimates and credible intervals from month $1$ to month $45$ with an increasing step size of $1$ month. We compare results under both BNPDM models with those given by \cite{tian2014predicting}'s method in Figure \ref{fig-chapter-3-real-data-2} and show a numerical summary in Table T1 (supplemental materials).

\section{Discussions \label{BNP-gSB-chap-discuss-future}}

In this article, we constructed a BNP estimation framework for estimating treatment effect measured by both average group differential RMST and subject-level RMST. \cite{zhang2022bayesian} proposed a BNP RMST estimator by putting mixture of Dirichlet process priors on the cumulative distribution function of the survival time random variable. Taking a different route, we treat the density of survival function hierarchically as a mixture of kernel densities where the mixtures have a (dependent) stick-breaking process prior. Our modeling approach is analogous to that of a dependent DP mixture (DDPM) model but with a more flexible stick-breaking probability assignment mechanism. A major advantage of our approach is to enable adjustments of mixed-type covariates/predictors. While many works on modeling spatial data focus on dependent structures that adjust for continuous covariates \citep{reich2007multivariate,ren2011logistic,diana2020hierarchical}, mixed-type covariates are more often seen in clinical settings. When the data generation model is a function of covariates, both group-level and subject-level RMST inference could be less efficient or inconsistent, without properly adjusting for the observed covariates.

We proposed a novel dependent stick-breaking process prior: the SPSBP (shrinkage probit stick-breaking process) prior, which is inspired by \cite{rodriguez2011nonparametric}'s (dependent) PSBP (probit stick-breaking process) prior. The SPSBP prior results in less variable estimates (e.g., narrower credible intervals) given a small or moderate sample size compared to the PSBP prior. This shrinkage effect is achieved through a more efficient stick-breaking probability assignment process that utilizes the sample first and second (central) moments of the empirical density function of the linearly transformed covariates. However, since the level of shrinking is controlled by sample variance and inversely proportional to the observed sample size, having a huge sample size could results in assigning the entire unit length towards the first few sticks, which results in a less discrete realization of the stick-breaking process. Fortunately, we did not experience such issue when modeling data up to $2,000$ observations and $10$ clusters. In fact, we found out through simulation studies that modeling with a smaller cluster size ($3$ or $5$) often result in better performance, compared to using say $10$ clusters, under a sample size between $200$ and $2,000$. 

Our simulation studies show decent performance by BNPDM models on group-level RMSTD inference giving ignorable biases and CPs up to $90\%$ for a $95\%$ nominal under Weibull and lognormal data generation models (RCT setting). In comparison, the two frequentist methods \citep{tian2014predicting,ambrogi2022analyzing} give consistent point estimates and CPs that attain the nominal level of $95\%$ under the same data generation settings (Weibull-RCT or lognormal-RCT). Credible sets of infinite-dimensional Bayesian models are not automatically frequentist confidence sets, and it is not automatically true that they contain the truth with the probability at least the credible level \citep{szabo2015frequentist}. The less efficiency of certain BNP models is well studied in the literature. For example, see \citep{cox1993analysis,diaconis1997bernstein}. However, we found that BNPDM models' subject-level RMST prediction results are better than those of frequentist methods \citep{tian2014predicting,ambrogi2022analyzing} when the underlying data generation model is a two-component mixture of Weibulls. Besides, we found our BNPDM models have more robust performances against cases where treatment assignment is confounded by observed covariates compared to \cite{tian2014predicting,ambrogi2022analyzing}'s methods. 

\section*{Appendix}

Assuming a Weibull kernel density (scale=$\bm{\psi}(\bm{W})^\prime\bm{\beta}$, shape=$\omega$), (\ref{subj-estimator-def-2}) has a closed form
\begin{align*}
\begin{split}
&\widehat{RMST}(t = \tau \; | \; \bm{w}_i) = E_{\bm{\theta}}\Bigg[\sum_{h=1}^L \Bigg\{ \pi_h\big(\bm{\psi}(\bm{w}_i)^\prime\bm{\alpha}_h\big) \cdot \Big[\tau \cdot S_{i,h}(\tau) + \bm{\psi}(\bm{w}_i)^\prime\bm{\beta}_h  \cdot \\ &  \gamma\big(\frac{1}{\omega_h}+1,\bm{\psi}(\bm{w}_i)^\prime\bm{\beta}_h^{(-\omega_h)} \tau^{\omega_h}\big) \Big] \Bigg\}\Bigg]
\end{split}
\end{align*}
where $\gamma(s,x)=\int_0^x t^{s-1} e^{-t} dt$ is the lower incomplete gamma function and $S_{i,h}(\tau) = exp\big\{-(\tau/\bm{\psi}(\bm{w}_i)^\prime\bm{\beta}_h)^{\omega_h}\big\}$. 

Assuming a Gamma kernel density (rate=$\bm{\psi}(\bm{W})^\prime\bm{\beta}$, shape=$\omega$), (\ref{subj-estimator-def-2}) has a closed form
\begin{align*}
\begin{split}
    &\widehat{RMST}(t = \tau \; | \; \bm{w}_i) = E_{\bm{\theta}}\Bigg[\sum_{h=1}^L \Bigg\{ \pi_h\big(\bm{\psi}(\bm{w}_i)^\prime\bm{\alpha}_h\big) \cdot \Big\{\tau + \\ & \Gamma(\omega_h)^{-1} \Big[ \frac{\omega_h \cdot \gamma\big(\omega_h, \bm{\psi}(\bm{w}_i)^\prime\bm{\beta}_h \tau\big) - \big(\bm{\psi}(\bm{w}_i)^\prime\bm{\beta}_h \tau\big)^{\omega_h} e^{-\big(\bm{\psi}(\bm{w}_i)^\prime\bm{\beta}_h \tau\big)}}{\bm{\psi}(\bm{w}_i)^\prime\bm{\beta}_h} - \tau \cdot \gamma\big(\omega_h, \bm{\psi}(\bm{w}_i)^\prime\bm{\beta}_h \cdot \tau\big)  \Big]\Big\} \Bigg\}\Bigg]
\end{split}
\end{align*}
given the property that $\gamma(s+1,x)=s \gamma(s,x) - x^s e^{-x}$ where $\gamma(s,x)=\int_0^x t^{s-1} e^{-t} dt$ is the lower incomplete gamma function and $\Gamma(\cdot)$ denotes the gamma function. 



\section*{Supplemental Materials}

\section*{Performance Evaluations}

\subsection*{Single-$\tau$ Group-Level Results and Subject-Level Results}

\begin{figure}[H]
 \centering
  \renewcommand{\baselinestretch}{1}
 \caption[Single time point analysis scenario S1]{Single time point analysis scenario S1: ATE point estimates with 95\% credible intervals comparing 4 prior models (Weibull data generation model under an RCT setting; $\tau=5$)}
     \begin{subfigure}[b]{0.45\textwidth}
      \includegraphics[width=\textwidth]{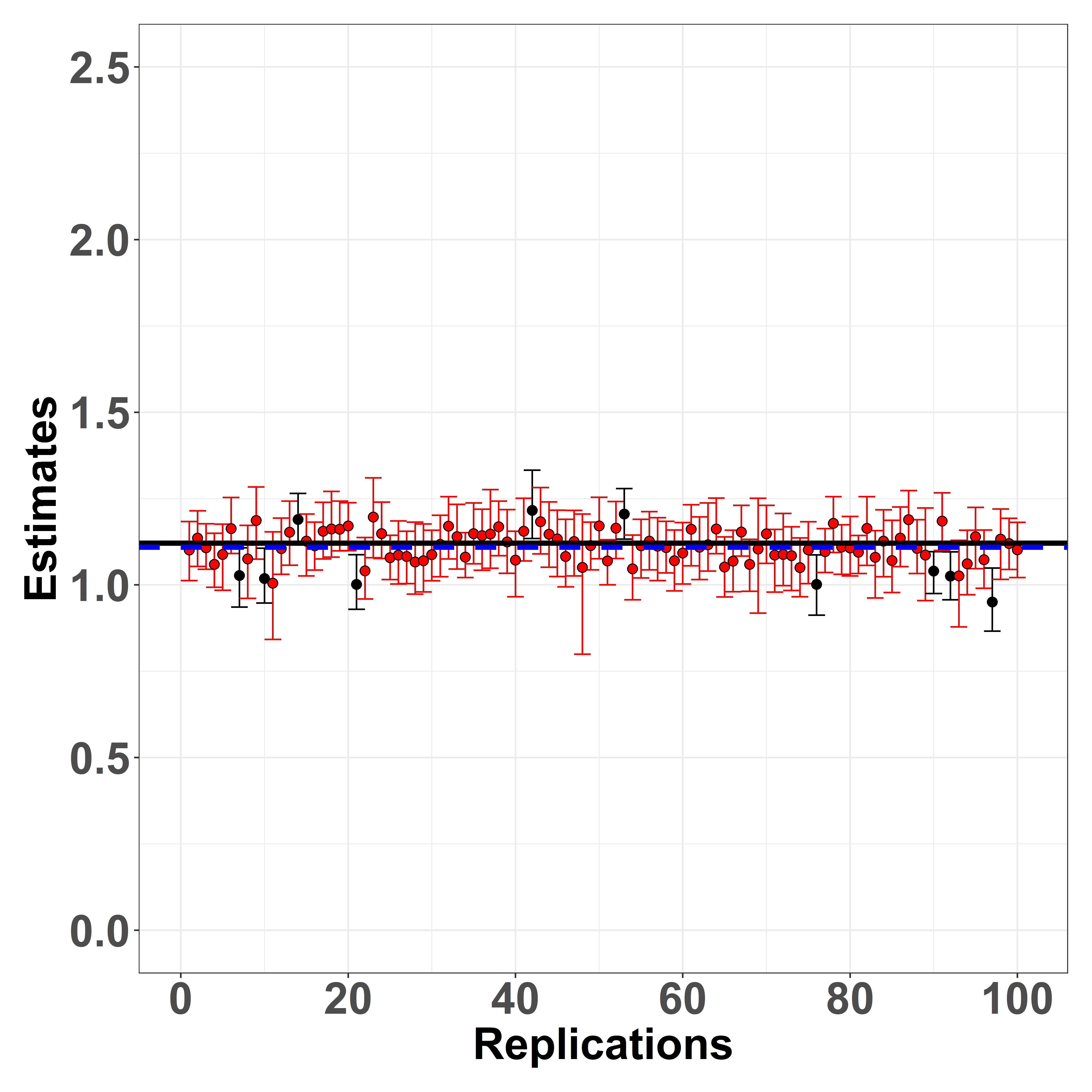}
      \caption{\centering SPSBP prior-Weibull kernel}
      \end{subfigure}
     \begin{subfigure}[b]{0.45\textwidth}
     \includegraphics[width=\textwidth]{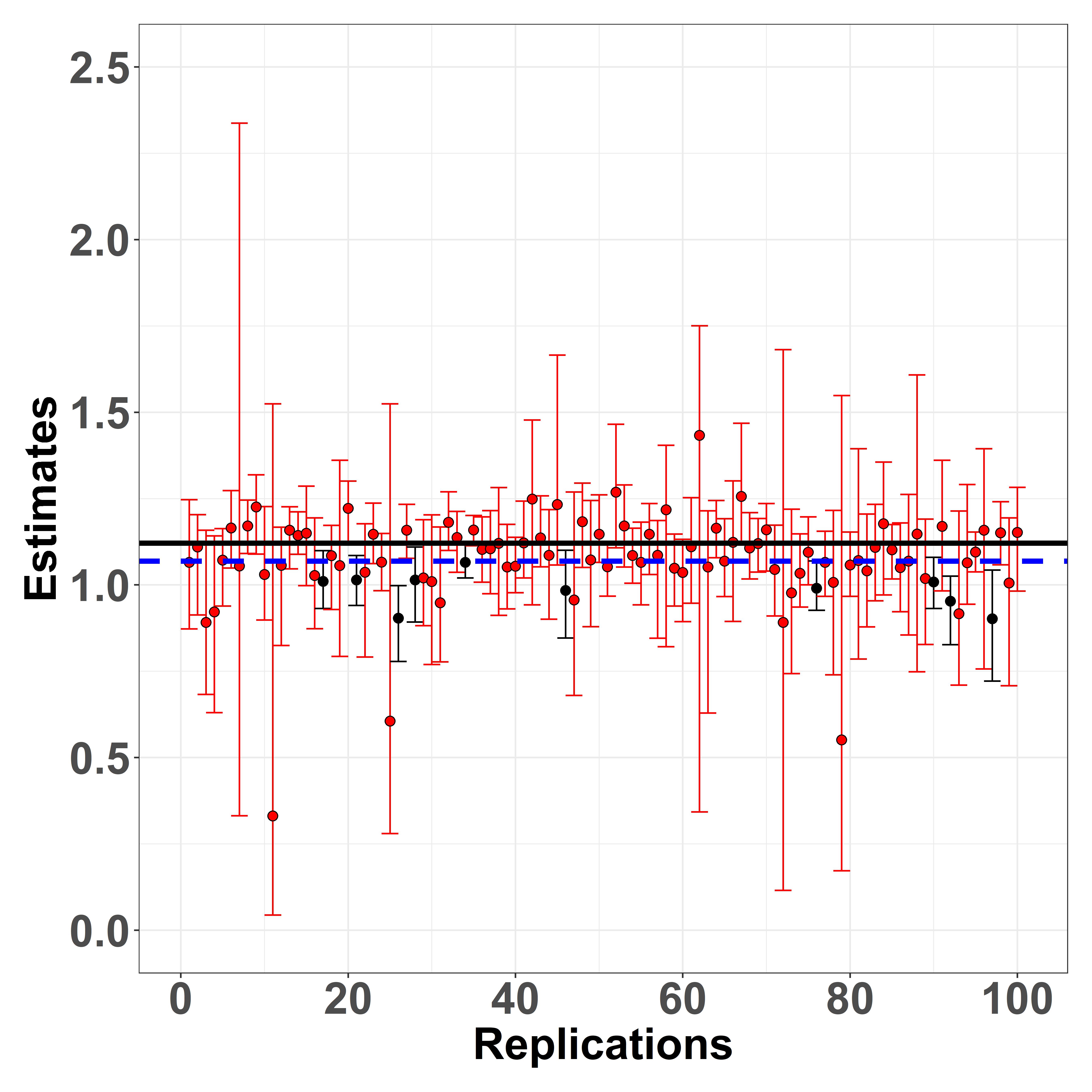}
      \caption{\centering PSBP prior-Weibull kernel}
      \end{subfigure}
          \begin{subfigure}[b]{0.45\textwidth}
      \includegraphics[width=\textwidth]{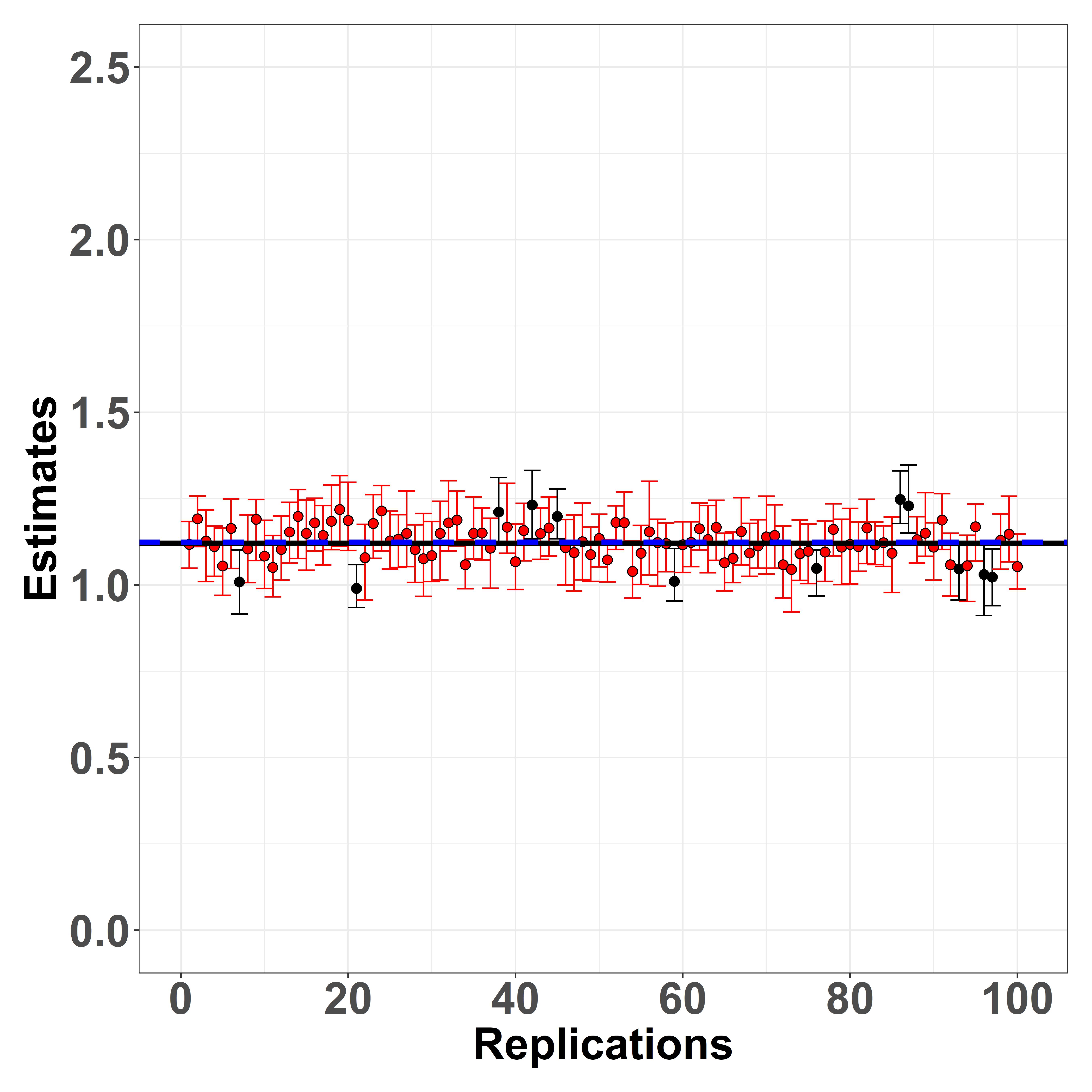}
     \caption{\centering LSBP prior-Weibull kernel}
    \end{subfigure}
               \begin{subfigure}[b]{0.45\textwidth}
      \includegraphics[width=\textwidth]{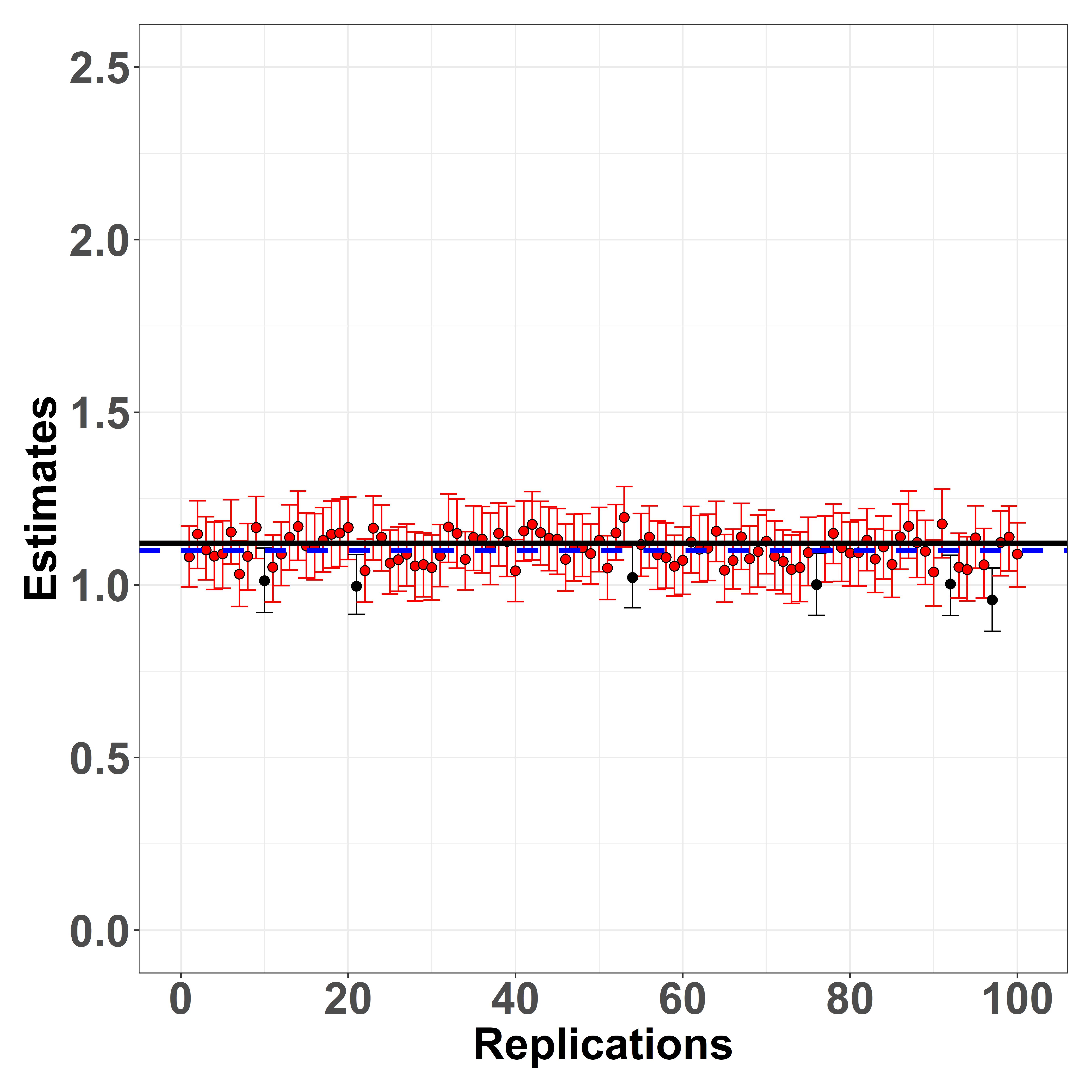}
      \caption{\centering LDDP prior-Weibull kernel}
      \end{subfigure}
     \caption*{
     RMSTD evaluated at $\tau=5$ years; coefficients of predictor effects: $(\beta_{A}=2.5,\beta_{W_1}=1.5,\beta_{W_2}=2.5,\beta_{W_3}=1.3)$; black solid lines and blue dashed lines mark the true RMSTD value and average RMSTD point estimates, respectively.}
     \label{fig-single-tau-ATE-F1}
 \end{figure}

Figure \ref{fig-single-tau-ATE-F1} shows results for group-level ATE inference under the Weibull data generation model (RCT setting), all four models (with Weibull kernel density) give unbiased estimates and satisfying CPs (from $88\%$ to $94\%$). Models with SPSBP prior, LSBP prior, and LDDP prior show similar properties and have narrower credible intervals compared to that of PSBP prior. The LDDP prior (model) shows the best results probably due to the fact that the analysis model (kernel density part) matches the data generation model. 

\begin{figure}[H]
\centering
\renewcommand{\baselinestretch}{1}
\caption[Single time point analysis scenario S1]{Single time point analysis scenario S1: ATE point estimates with 95\% credible intervals comparing 4 prior models (2-components Weibull data generation model under an RCT setting; $\tau=5$)}
     \begin{subfigure}[b]{0.45\textwidth}
     \includegraphics[width=\textwidth]{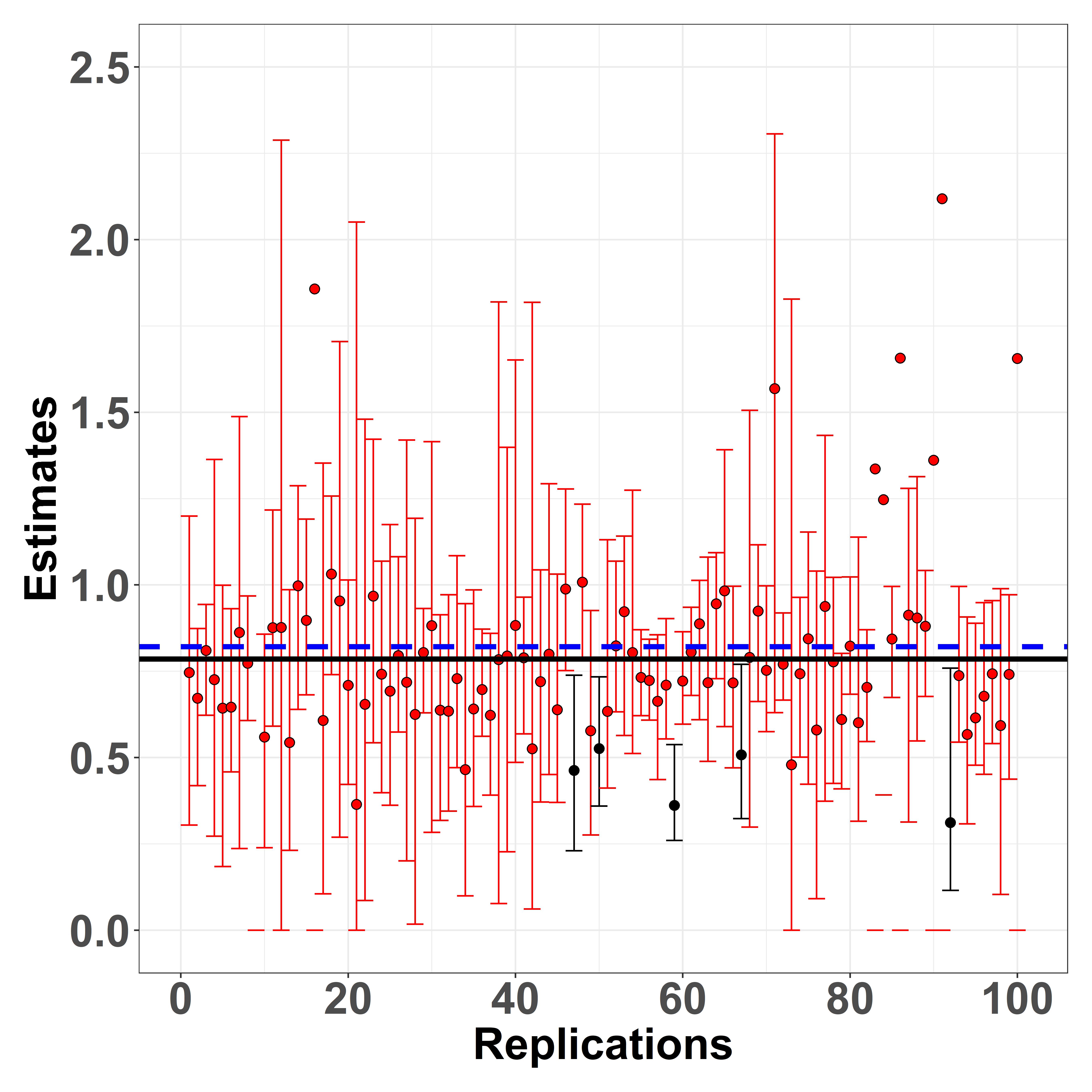}
      \caption{\centering SPSBP prior-Weibull kernel}
      \end{subfigure}
     \begin{subfigure}[b]{0.45\textwidth}
      \includegraphics[width=\textwidth]{figure-causal-2weibRCT-PSBPWeib-SS500-cluster5-warm2k-iter1k.jpeg}
     \caption{\centering PSBP prior-Weibull kernel}
     \end{subfigure}
         \begin{subfigure}[b]{0.45\textwidth}
    \includegraphics[width=\textwidth]{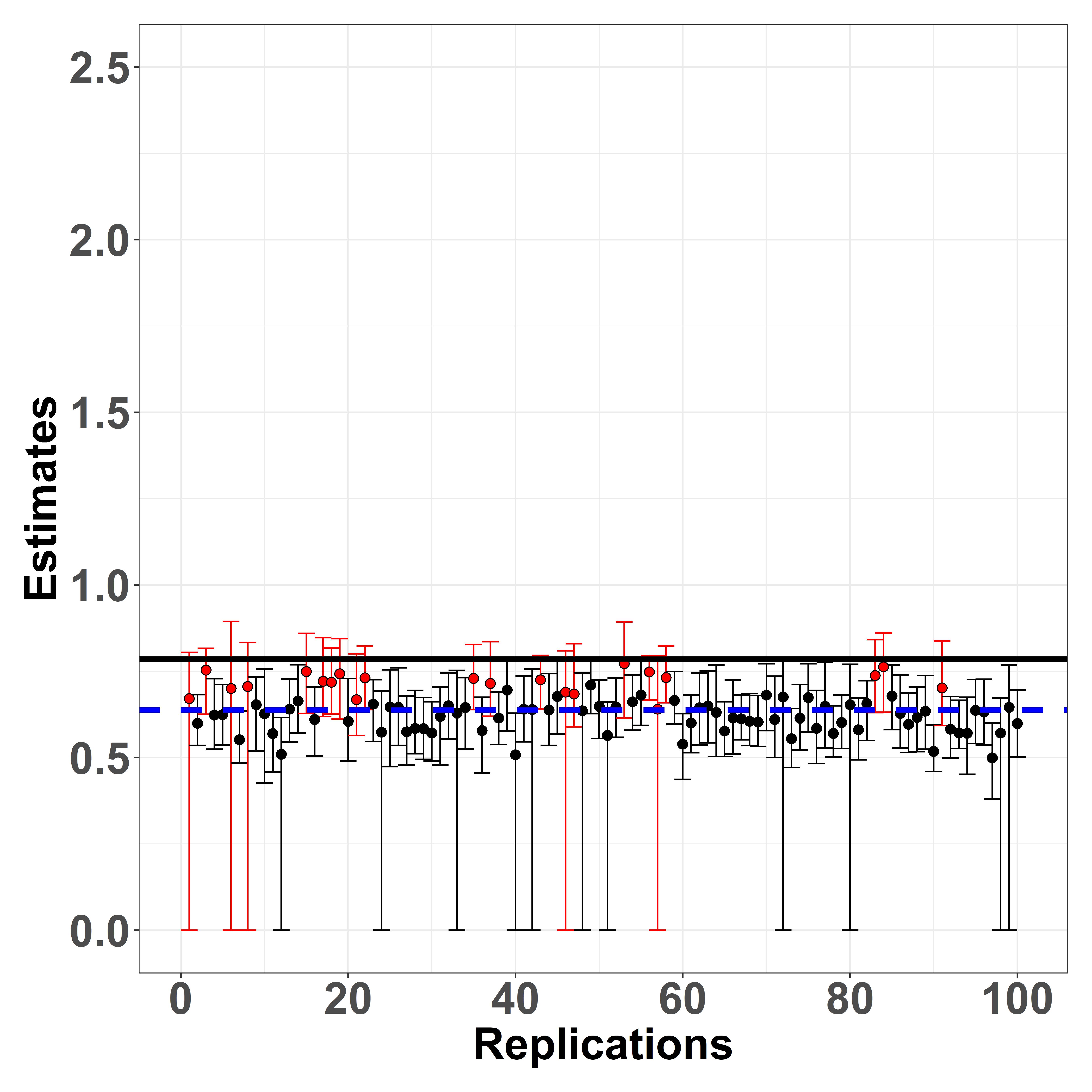}
      \caption{\centering LSBP prior-Weibull kernel}
     \end{subfigure}
              \begin{subfigure}[b]{0.45\textwidth}
      \includegraphics[width=\textwidth]{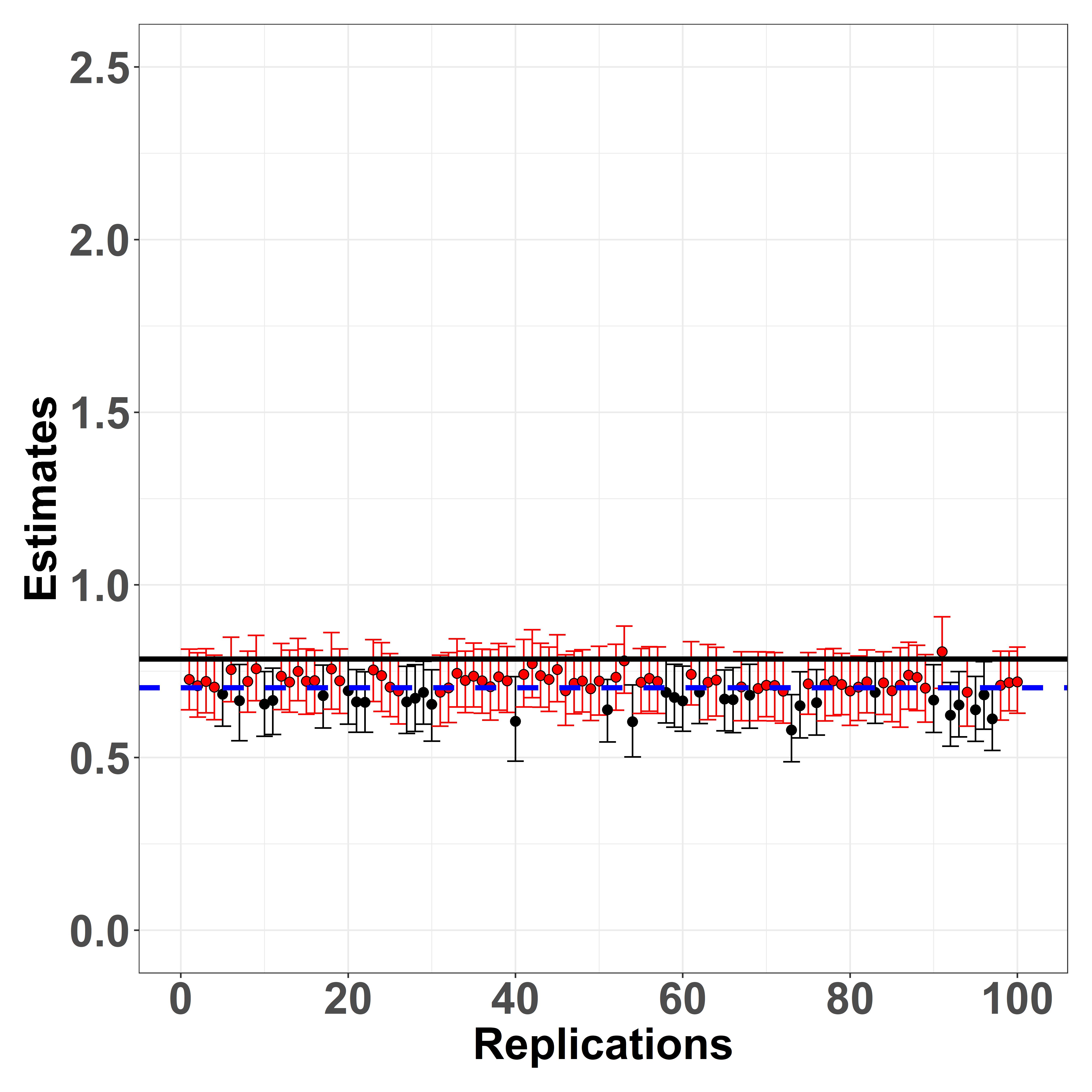}
     \caption{\centering LDDP prior-Weibull kernel}
      \end{subfigure}
     \caption*{
     RMSTD evaluated at $\tau=5$ years; coefficients of predictor effects: $(\beta_{A}=2.5,\beta_{W_1}=1.5,\beta_{W_2}=2.5,\beta_{W_3}=1.3)$; black solid lines and blue dashed lines mark the true RMSTD value and average RMSTD point estimates, respectively.}
     \label{fig-single-tau-ATE-F2}
\end{figure}

Figure \ref{fig-single-tau-ATE-F2} shows results under the two-components Weibull mixture data generation model. The LSBP prior shows unstable results with several credible interval lower bounds down to zero. The PSBP prior still has volatile performance with possible signs of divergence. All priors (except for the PSBP prior) show small to moderate biases (from $-0.08$ to $-0.15$). 

\begin{figure}[H]
  \vspace{-1em}
 \centering
  \renewcommand{\baselinestretch}{1}
      \caption[Subject level RMST predictions scenario 1-B]{\small Subject level RMST predictions scenario 1-B: subject versus bias; data are generated under the two-components Weibull mixture model; $\tau=5$}
 \includegraphics[scale=0.45]{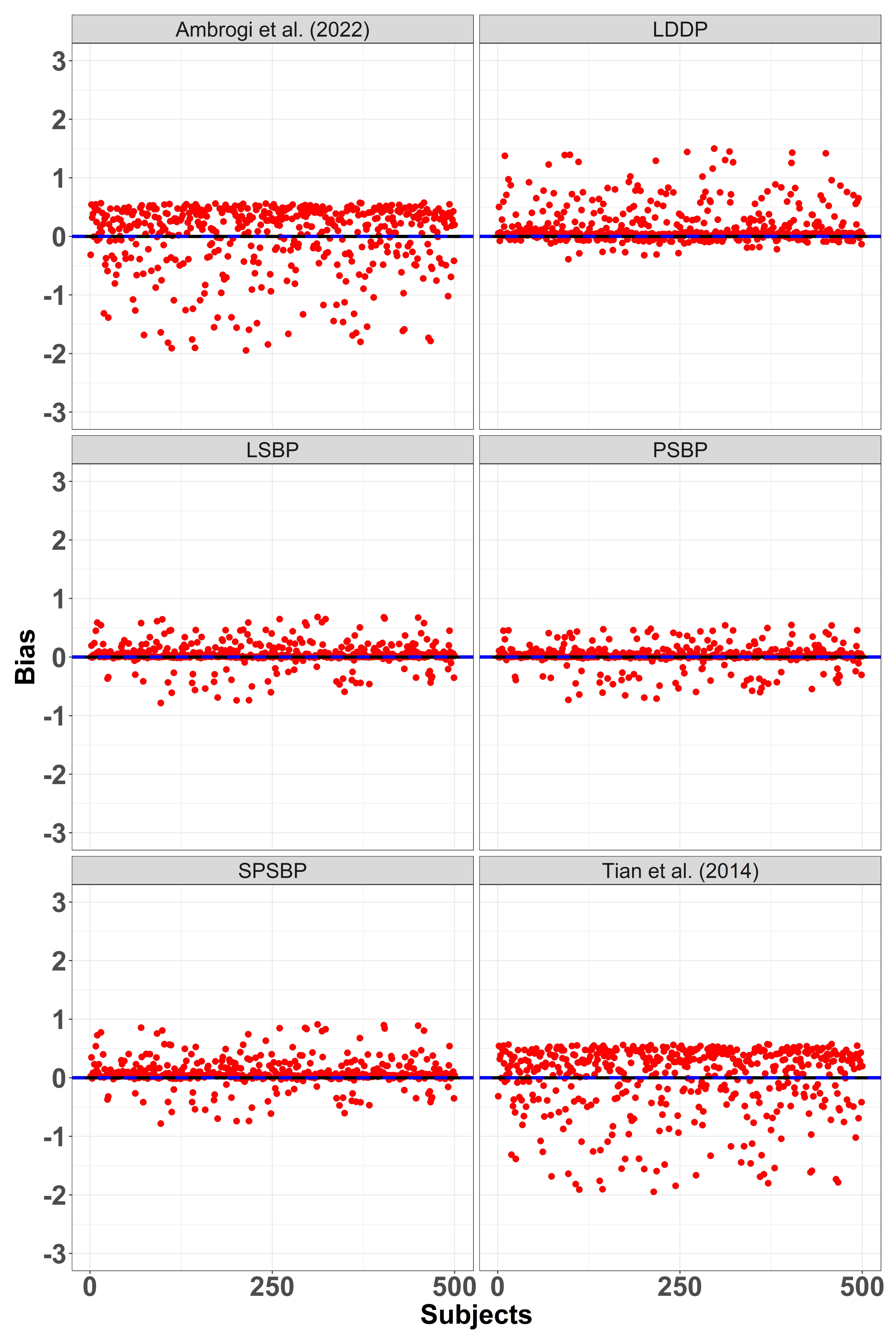}
     \label{fig-chap3-subjs-plot-1}
 \end{figure}

Figure \ref{fig-chap3-subjs-plot-1} is a (points) scatter plot showing individual-level RMST prediction bias for each subject, under the two-components Weibull data generation model with various approaches. The two frequentist methods \citep{tian2014predicting,ambrogi2022analyzing} result in large negatively biased estimates and dense small positively biased estimates on average. The LDDP prior's results are partially positively biased but mostly condensed on the zero bias line. In contrast, the results given by three DSBP prior models have the least biases and RMSEs (See Table 1 in the manuscript). For all three models (SPSBP, PSBP, and LSBP), the bias scatters are narrowly and evenly distributed around the zero bias line.

 \begin{figure}[H]
 \centering
  \renewcommand{\baselinestretch}{1}
      \caption[Subject level RMST predictions scenario 1-C]{\small Subject level RMST predictions scenario 1-C: $\bm{w}_i^{tr} \beta_{2Weibull} (i=\{1,\ldots,N\})$ versus bias; data are generated under the two-components Weibull mixture model; $\tau=5$}
 \includegraphics[scale=0.45]{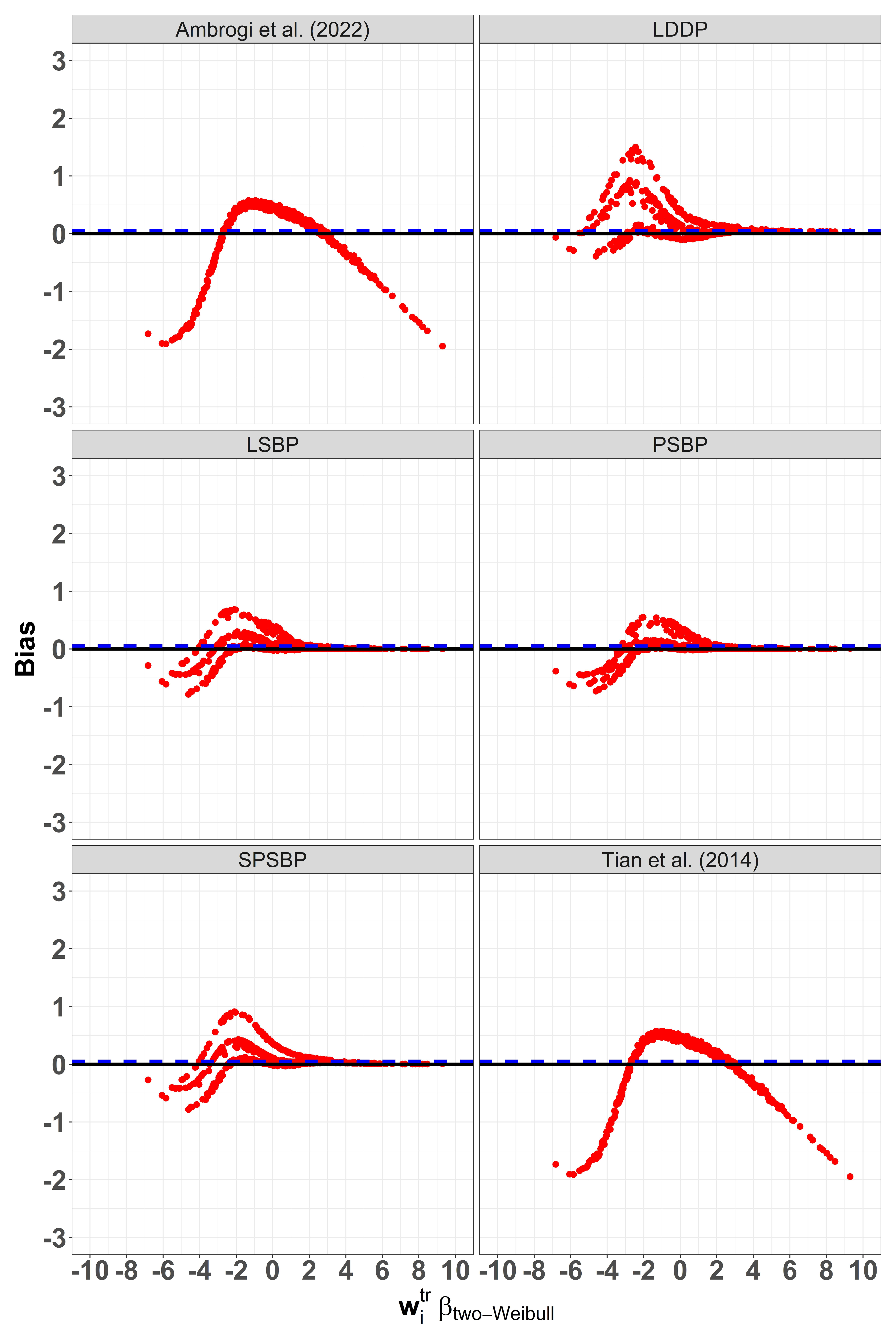}
     \label{fig-chap3-subjs-plot-2}
 \end{figure}

\subsection*{Real Data Analysis Results}

\begin{figure}[H]
\centering
 \renewcommand{\baselinestretch}{1}
\caption{Estimated survival probabilities}
    \begin{subfigure}[b]{\textwidth}
     \includegraphics[width=\textwidth]{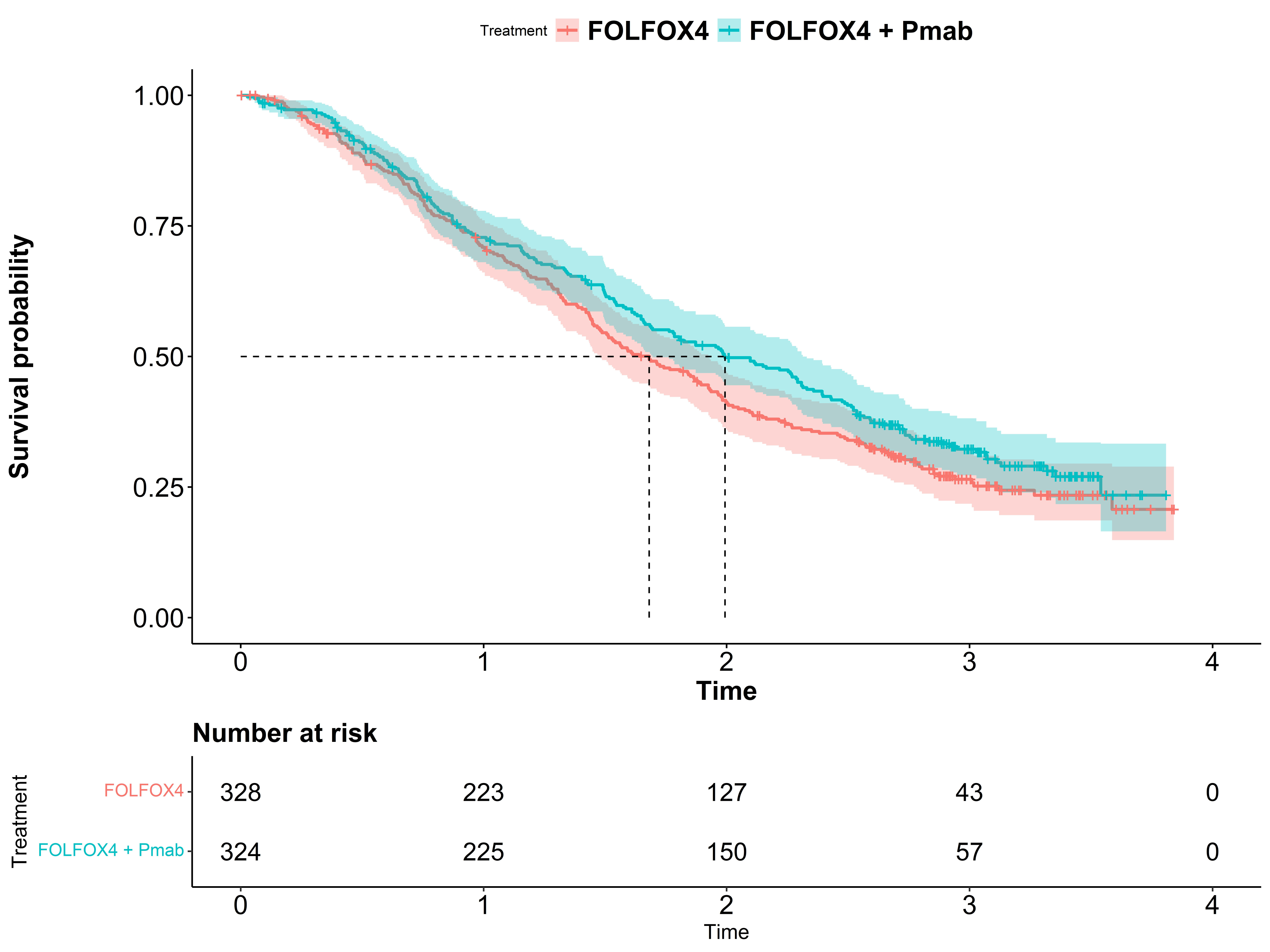}
     \end{subfigure}
    \caption*{\small A Cox proportional hazards model \citep{therneau2000cox} that adjusts for bmi and kras is fitted to estimate survival probabilities; both point estimates (colored solid lines) and $95\%$ confidence intervals (colored shades) are presented; median survival time is marked by black dotted lines.}
    \label{fig-chapter-3-real-data-1}
\end{figure}

Figure \ref{fig-chapter-3-real-data-1} shows survival probabilities estimated for the two treatment groups of PRIME trial data on the OS endpoint among KRAS WT patients. A Cox proportional hazards model \citep{therneau2000cox} that adjusts for body mass index (BMI) and age was fitted. Both point estimates (colored solid lines) and $95\%$ confidence intervals (colored shades) are presented, and median survival time is marked by black dotted lines.

\begin{table}[H]
\centering
\caption[REAL DATA RMSTD CURVE ANALYSIS]{REAL DATA RMSTD CURVE ANALYSIS
}
\vspace{5pt}
\renewcommand{\baselinestretch}{1}
    \setlength{\tabcolsep}{6pt} 
\renewcommand{\arraystretch}{1}
\resizebox{.8\textwidth}{!}{%
\begin{tabular}{|c|c|c|c|c|c|c|c|c|}
\hline
\textbf{Method} & \textbf{$\tau$ (years)} & \textbf{0.5} & \textbf{1} & \textbf{1.5} & \textbf{2} & \textbf{2.5} & \textbf{3} & \textbf{3.5} \\ \hline
\multirow{4}{*}{\textbf{\begin{tabular}[c]{@{}c@{}}LSBP\\ Weibull\end{tabular}}} & \textit{\begin{tabular}[c]{@{}c@{}}$95\%$ \\ CI Lower\end{tabular}} & 0.00 & 0.00 & 0.00 & 0.00 & 0.00 & 0.00 & -0.01 \\ \cline{2-9} 
 & \textit{\begin{tabular}[c]{@{}c@{}}Point \\ Estimate\end{tabular}} & 0.01 & 0.02 & 0.04 & 0.08 & 0.11 & 0.14 & 0.15 \\ \cline{2-9} 
 & \textit{\begin{tabular}[c]{@{}c@{}}$95\%$ \\ CI Upper\end{tabular}} & 0.01 & 0.04 & 0.09 & 0.16 & 0.22 & 0.27 & 0.31 \\ \hline
\multirow{4}{*}{\textbf{\begin{tabular}[c]{@{}c@{}}SPSBP\\ Weibull\end{tabular}}} & \textit{\begin{tabular}[c]{@{}c@{}}$95\%$ \\ CI Lower\end{tabular}} & 0.00 & -0.01 & -0.02 & -0.03 & -0.04 & -0.06 & -0.07 \\ \cline{2-9} 
 & \textit{\begin{tabular}[c]{@{}c@{}}Point \\ Estimate\end{tabular}} & 0.00 & 0.02 & 0.04 & 0.06 & 0.09 & 0.11 & 0.12 \\ \cline{2-9} 
 & \textit{\begin{tabular}[c]{@{}c@{}}$95\%$ \\ CI Upper\end{tabular}} & 0.01 & 0.04 & 0.10 & 0.16 & 0.22 & 0.26 & 0.30 \\ \hline
 \multirow{4}{*}{\textbf{\begin{tabular}[c]{@{}c@{}}PSBP\\ Weibull\end{tabular}}} & \textit{\begin{tabular}[c]{@{}c@{}}$95\%$ \\ CI Lower\end{tabular}} & 0.00 & 0.00 & 0.00 & 0.00 & 0.00 & 0.00 & 0.00 \\ \cline{2-9} 
 & \textit{\begin{tabular}[c]{@{}c@{}}Point \\ Estimate\end{tabular}} & 0.01 & 0.03 & 0.07 & 0.13 & 0.19 & 0.24 & 0.29 \\ \cline{2-9} 
 & \textit{\begin{tabular}[c]{@{}c@{}}$95\%$ \\ CI Upper\end{tabular}} & 0.02 & 0.07 & 0.15 & 0.26 & 0.38 & 0.49 & 0.59 \\ \hline
 \multirow{4}{*}{\textbf{\begin{tabular}[c]{@{}c@{}}LSBP\\ Gamma\end{tabular}}} & \textit{\begin{tabular}[c]{@{}c@{}}$95\%$ \\ CI Lower\end{tabular}} & 0.00 & 0.00 & 0.00 & 0.00 & 0.00 & 0.00 & 0.00 \\ \cline{2-9} 
 & \textit{\begin{tabular}[c]{@{}c@{}}Point \\ Estimate\end{tabular}} & 0.00 & 0.02 & 0.04 & 0.07 & 0.11 & 0.14 & 0.17 \\ \cline{2-9} 
 & \textit{\begin{tabular}[c]{@{}c@{}}$95\%$ \\ CI Upper\end{tabular}} & 0.01 & 0.04 & 0.09 & 0.15 & 0.22 & 0.28 & 0.34 \\ \hline
\multirow{4}{*}{\textbf{\begin{tabular}[c]{@{}c@{}}SPSBP\\ Gamma\end{tabular}}} & \textit{\begin{tabular}[c]{@{}c@{}}$95\%$ \\ CI Lower\end{tabular}} & 0.00 & -0.01 & -0.01 & -0.02 & -0.02 & -0.03 & -0.03 \\ \cline{2-9} 
 & \textit{\begin{tabular}[c]{@{}c@{}}Point \\ Estimate\end{tabular}} & 0.00 & 0.02 & 0.04 & 0.07 & 0.10 & 0.13 & 0.15 \\ \cline{2-9} 
 & \textit{\begin{tabular}[c]{@{}c@{}}$95\%$ \\ CI Upper\end{tabular}} & 0.01 & 0.05 & 0.10 & 0.16 & 0.23 & 0.30 & 0.37 \\ \hline
 \multirow{4}{*}{\textbf{\begin{tabular}[c]{@{}c@{}}PSBP\\ Gamma\end{tabular}}} & \textit{\begin{tabular}[c]{@{}c@{}}$95\%$ \\ CI Lower\end{tabular}} & -0.01 & -0.03 & -0.06 & -0.10 & -0.15 & -0.20 & -0.25 \\ \cline{2-9} 
 & \textit{\begin{tabular}[c]{@{}c@{}}Point \\ Estimate\end{tabular}} & 0.00 & 0.01 & 0.03 & 0.05 & 0.07 & 0.09 & 0.11 \\ \cline{2-9} 
 & \textit{\begin{tabular}[c]{@{}c@{}}$95\%$ \\ CI Upper\end{tabular}} & 0.01 & 0.05 & 0.10 & 0.17 & 0.24 & 0.31 & 0.38 \\ \hline
\multirow{4}{*}{\textbf{\cite{tian2014predicting}}} & \textit{\begin{tabular}[c]{@{}c@{}}$95\%$ \\ CI Lower\end{tabular}} & -0.01 & -0.02 & -0.03 & -0.03 & -0.02 & -0.01 & -0.04 \\ \cline{2-9} 
 & \textit{\begin{tabular}[c]{@{}c@{}}Point \\ Estimate\end{tabular}} & 0.01 & 0.01 & 0.03 & 0.07 & 0.11 & 0.14 & 0.14 \\ \cline{2-9} 
 & \textit{\begin{tabular}[c]{@{}c@{}}$95\%$ \\ CI Upper\end{tabular}} & 0.02 & 0.05 & 0.10 & 0.16 & 0.24 & 0.30 & 0.33 \\ \hline
\end{tabular}}
          \label{tab-chap3-real-data-table1}
\end{table}

\section*{Additional Results}

\subsection*{Linear Dependent Dirichlet Process Mixture Models \label{chap3-posterior-scheme-LDDP}}






We can model a density function through a Dirichlet process mixture (DPM) model: $f(y) = \int K_{\bm{\theta}}(y) \; dG(\bm{\theta})$ where $G \sim DP(\alpha,G_0)$. 

Alternatively, we can incorporate covariates dependence on (certain) parameter(s) of the kernel density, which results in a LDDP mixture model:
\vspace{-.5em}
\begin{equation}
    f(y) = \int K_{}\big(y \; | \; \bm{\theta}(\bm{w})\big) \; dG_{}\big(\bm{\theta}(\bm{w})\big); \quad G_{} \sim DP(\alpha,G_0)
\end{equation}

where $\bm{\theta}(\bm{w})=\big(\bm{\psi}(\bm{w})^T \bm{\beta},\bm{\omega}\big)$. The DP prior is assigned on $\bm{\beta}$ and $\bm{\omega}$: $(\bm{\beta},\bm{\omega}) \sim G$ where $\bm{\beta}=\{\bm{\beta}_1,\ldots,\bm{\beta}_R\}$ for adjusting $R$ predictors. 
In the above formulation, covariates dependence are only introduced on the point masses (atoms), which categorize it as a single-$\pi$ DDP model \cite{quintana2020dependent}. 
For the $h^{th}$ cluster, 

\vspace{-.5em}
\begin{align}
\begin{split}
    \label{LDDP-mix-mod-spec-1}
    &y_i \ | \ \bm{\beta}_{i,h}, \omega_h \stackrel{i.i.d.}{\sim} K\big(y_i \; | \; exp\big\{\bm{\psi}(\bm{w}_i)^{tr}\bm{\beta}_{i,h}\big\}, \omega_{h}\big), \quad i=1,\ldots,N \\
    &v_h \sim Beta(1,M), \; \pi_1=v_1 \\  &\pi_h=\big(1-v_1\big)\big(1-v_2\big)\ldots\big(1-v_{h-1}\big)v_h, \quad h \in\{2,\ldots,L-1\} \\
        &\bm{\beta}_h=\{\beta_{h,A},\beta_{h,1},\ldots,\beta_{h,R}\} \sim \mathcal{N}_{R+1}(\bm{\mu}_\beta,\bm{\Sigma}_\beta), \quad h \in \{1,\ldots,L\} 
        \\ &\omega_h \sim unif(c_{low},c_{up}), \quad h \in \{1,\ldots,L\}
\end{split}
\end{align}

where $(M,\bm{\mu}_\beta,\bm{\Sigma}_\beta,c_{low},c_{up})$ are constants. For more flexibility, one could model $M \sim Gamma(a_{\alpha},b_{\alpha})$ given constants $(a_{\alpha},b_{\alpha})$. As previously stated, we can choose $K(y_i \; | \; \cdot)$ to be either a Weibull kernel density (scale$=exp\big\{\bm{\psi}(\bm{w}_i)^{tr}\bm{\beta}_{ih}\big\}$, shape$=\omega_{h}$) or a gamma kernel density (rate$=exp\big\{\bm{\psi}(\bm{w}_i)^{tr}\bm{\beta}_{ih}\big\}$, shape$=\omega_{h}$) whose support is on $\mathbb{R}^+$.


Subsequently, its log joint density is
\vspace{-.5em}
\begin{align}
\begin{split}
    &\mathcal{Q}\big(\bm{\theta}=(\bm{\beta},\bm{\omega},M),\bm{o}\big) = \prod_{i=1}^N \Big\{\big[f_{\bm{\theta}}(y_i)\big(1-H(y_i)\big)\big]^{\delta_i}\big[S_{\bm{\theta}}(y_i) h(y_i)\big]^{(1-\delta_i)}\Big\} \\
    & \propto \prod_{i=1}^N \big[ f_{\bm{\theta}}(y_i)^{\delta_i} S_{\bm{\theta}}(y_i)^{(1-\delta_i)} \big] \\
    &\propto \prod_{i=1}^N \sum_{j=1}^{L} \pi_{j} \cdot \Big[ K\big(y_i \; | \; exp\big\{\bm{\psi}(\bm{w}_i)^{tr}\bm{\beta}_{i,j}\big\}, \omega_{j}\big) \Big]^{\delta_i} \cdot \Big[ \int_t^\infty K\big(y_i \; | \; exp\big\{\bm{\psi}(\bm{w}_i)^{tr}\bm{\beta}_{i,j}\big\}, \omega_{j}\big) \Big]^{(1-\delta_i)} \\
    &\propto \prod_{i=1}^N \sum_{j=1}^{L} \Big\{v_j \prod_{l < j}\big[1-v_l\big]\Big\} \\ &
    \cdot \Big[ K\big(y_i \; | \; exp\big\{\bm{\psi}(\bm{w}_i)^{tr}\bm{\beta}_{i,j}\big\}, \omega_{j}\big) \Big]^{\delta_i} \cdot \Big[ \int_t^\infty K\big(y_i \; | \; exp\big\{\bm{\psi}(\bm{w}_i)^{tr}\bm{\beta}_{i,j}\big\}, \omega_{j}\big) \Big]^{(1-\delta_i)} 
\end{split}
\end{align}
where $v_j \sim Beta(1,M)$, $\bm{\beta}=\{\bm{\beta}_{i,j} \; | \; i=1,\ldots,N,j=1,\ldots,L\}; \; \bm{\beta}_{i,j}=\{\bm{\beta}_{i,j,1},\ldots,\bm{\beta}_{i,j,R}\}$, $\bm{\omega}=\{\omega_1,\ldots,\omega_L\}$, $\bm{o}=\{\bm{o}_1,\ldots,\bm{o}_N\}; \; o_i=\{y_i,\delta_i,\bm{w}_i\}$.

\subsection*{Single-Atoms Dependent Stick-Breaking Process Prior Mixture Models \label{chap3-posterior-scheme-single-atoms}}
We can model the RMST function by assigning a single-atoms DSBP prior
. This modeling approach assumes predictor dependence only on the mixing probabilities, which resembles a single-atoms dependent Dirichlet process prior mixture model. With a slight abuse of notations, let $\bm{\omega}=\{\omega_1,\omega_2\}$ denote the parameters of a two-parameter kernel density
. For this approach, we model both parameters $\{\omega_1,\omega_2\}$ assuming a (generalized) stick-breaking process prior, without incorporating covariates-dependence on the kernel densities. 

For the $h^{th}$ cluster, 
\vspace{-.5em}
\begin{align}
    \begin{split}
    \label{gSB-single-atoms-mix-mod-spec-1}
    &y_i \ | \ \omega_{1,h},\omega_{2,h} \stackrel{i.i.d.}{\sim} K\big(y_i \; | \; \omega_{1,h},\omega_{2,h} \big), \quad i=1,\ldots,N \\
    &v_h(\bm{w}_i) = g(\bm{\psi}(\bm{w}_i)^{tr} \alpha_h), \; \pi_1(\bm{w}_i)=v_1(\bm{w}_i) \\  &\pi_h(\bm{w}_i)=\big(1-v_1(\bm{w}_i)\big)\big(1-v_2(\bm{w}_i)\big)\ldots\big(1-v_{h-1}(\bm{w}_i)\big)v_h(\bm{w}_i), \quad h \in\{2,\ldots,L-1\} \\
        &\bm{\alpha}_h=\{\alpha_{h,A},\alpha_{h,1},\ldots,\alpha_{h,R}\} \sim \mathcal{N}_{R+1}(\bm{\mu}_\alpha,\bm{\Sigma}_\alpha), \quad h \in \{1,\ldots,L-1\} 
        \\ &\omega_{1,h} \sim unif(c_{low},c_{up}), \; \omega_{2,h} \sim unif(d_{low},d_{up}), \quad h \in \{1,\ldots,L\}
    \end{split}
\end{align}

where $(\bm{\mu}_\alpha,\bm{\Sigma}_\alpha,c_{low},c_{up},d_{low},d_{up})$ are constants. For more flexibility, one could model $\omega_1 \sim Gamma(a_{1},b_{1}); \; \omega_2 \sim Gamma(a_{2},b_{2})$ given constants $(a_{1},b_{1},a_{2},b_{2})$. The log joint density is
\vspace{-.5em}
\begin{align}
\begin{split}
    &\mathcal{Q}\big(\bm{\theta}=(\bm{\alpha},\bm{\omega}_1,\bm{\omega}_2),\bm{o}\big) = \prod_{i=1}^N \Big\{\big[f_{\bm{\theta}}(y_i)\big(1-H(y_i)\big)\big]^{\delta_i}\big[S_{\bm{\theta}}(y_i) h(y_i)\big]^{(1-\delta_i)}\Big\} \\ &\propto \prod_{i=1}^N \big[ f_{\bm{\theta}}(y_i)^{\delta_i} S_{\bm{\theta}}(y_i)^{(1-\delta_i)} \big] \\
    &\propto \prod_{i=1}^N \sum_{j=1}^{L} \pi_{j}(\bm{w}_i) \cdot \Big[ K\big(y_i \; | \; \omega_{1,j}, \omega_{2,j}\big) \Big]^{\delta_i} \cdot \Big[ \int_t^\infty K\big(y_i \; | \; \omega_{1,j}, \omega_{2,j}\big) \Big]^{(1-\delta_i)} \\ &\propto \prod_{i=1}^N \sum_{j=1}^{L} \Big\{v_j(\bm{w}_i) \prod_{l < j}\big[1-v_l(\bm{w}_i)\big]\Big\} \cdot \Big[ K\big(y_i \; | \; \omega_{1,j}, \omega_{2,j}\big) \Big]^{\delta_i} \cdot \Big[ \int_t^\infty K\big(y_i \; | \; \omega_{1,j}, \omega_{2,j}\big) \Big]^{(1-\delta_i)} \\
    &\propto \prod_{i=1}^N \sum_{j=1}^{L} \Big\{g(\bm{\psi}(\bm{w}_i)^{tr} \bm{\alpha}_{i,j}) \prod_{l < j}\big[1-g(\bm{\psi}(\bm{w}_i)^{tr} \bm{\alpha}_{i,j})\big] \Big\} \cdot \Big[ K\big(y_i \; | \; \omega_{1,j}, \omega_{2,j}\big) \Big]^{\delta_i} \\& \cdot \Big[ \int_t^\infty K\big(y_i \; | \; \omega_{1,j}, \omega_{2,j}\big) \Big]^{(1-\delta_i)}
\end{split}
\end{align} 

where $\bm{\alpha}=\{\bm{\alpha}_{i,j} \; | \; i=1,\ldots,N,j=1,\ldots,L-1\}; \; \bm{\alpha}_{i,j}=\{\bm{\alpha}_{i,j,1},\ldots,\bm{\alpha}_{i,j,R}\}$, $\bm{\omega}=\{\bm{\omega}_1,\bm{\omega}_2\}; \; \bm{\omega}_1=\{\omega_{1,1},\ldots,\omega_{1,L}\}; \; \bm{\omega}_2=\{\omega_{2,1},\ldots,\omega_{2,L}\}$, $\bm{o}=\{\bm{o}_1,\ldots,\bm{o}_N\}; \; o_i=\{y_i,\delta_i,\bm{w}_i\}$.

Assuming a Weibull kernel density (scale=$\bm{\psi}(\bm{W})^\prime\bm{\beta}$, shape=$\omega$), (\ref{subj-estimator-def-2}) has a closed form
\begin{align*}
\begin{split}
&\widehat{RMST}(t = \tau \; | \; \bm{w}_i) = E_{\bm{\theta}}\Bigg[\sum_{h=1}^L \Bigg\{ \pi_h\big(\bm{\psi}(\bm{w}_i)^\prime\bm{\alpha}_h\big) \cdot \Big[\tau \cdot S_{i,h}(\tau) + \bm{\psi}(\bm{w}_i)^\prime\bm{\beta}_h  \cdot \\ &  \gamma\big(\frac{1}{\omega_h}+1,\bm{\psi}(\bm{w}_i)^\prime\bm{\beta}_h^{(-\omega_h)} \tau^{\omega_h}\big) \Big] \Bigg\}\Bigg]
\end{split}
\end{align*}

where $\gamma(s,x)=\int_0^x t^{s-1} e^{-t} dt$ is the lower incomplete gamma function and $S_{i,h}(\tau) = exp\big\{-(\tau/\bm{\psi}(\bm{w}_i)^\prime\bm{\beta}_h)^{\omega_h}\big\}$. 

Assuming a Gamma kernel density (rate=$\bm{\psi}(\bm{W})^\prime\bm{\beta}$, shape=$\omega$), (\ref{subj-estimator-def-2}) has a closed form
\begin{align*}
\begin{split}
    &\widehat{RMST}(t = \tau \; | \; \bm{w}_i) = E_{\bm{\theta}}\Bigg[\sum_{h=1}^L \Bigg\{ \pi_h\big(\bm{\psi}(\bm{w}_i)^\prime\bm{\alpha}_h\big) \cdot \Big\{\tau + \\ & \Gamma(\omega_h)^{-1} \Big[ \frac{\omega_h \cdot \gamma\big(\omega_h, \bm{\psi}(\bm{w}_i)^\prime\bm{\beta}_h \tau\big) - \big(\bm{\psi}(\bm{w}_i)^\prime\bm{\beta}_h \tau\big)^{\omega_h} e^{-\big(\bm{\psi}(\bm{w}_i)^\prime\bm{\beta}_h \tau\big)}}{\bm{\psi}(\bm{w}_i)^\prime\bm{\beta}_h} - \tau \cdot \gamma\big(\omega_h, \bm{\psi}(\bm{w}_i)^\prime\bm{\beta}_h \cdot \tau\big)  \Big]\Big\} \Bigg\}\Bigg]
\end{split}
\end{align*}
given the property that $\gamma(s+1,x)=s \gamma(s,x) - x^s e^{-x}$ where $\gamma(s,x)=\int_0^x t^{s-1} e^{-t} dt$ is the lower incomplete gamma function and $\Gamma(\cdot)$ denotes the gamma function. 

Define the causal RMSTD estimand and a BNP estimator as follows:

\begin{align*}
    \Delta(\tau) &= \text{RMST}_{A_1}(t=\tau) - \text{RMST}_{A_0}(t=\tau) \\
    &=\int_0^\tau E_{\bm{X}}\Big[S_{\bm{\theta}}(t \; | \;A=1, \bm{X})\Big] -  E_{\bm{X}}\Big[S_{\bm{\theta}}(t \; | \;A=0, \bm{X})\Big] \; dt 
\end{align*}


\begin{align}
    \begin{split}
    \label{causal-estimator-BNP-def-1}
        \widehat{\Delta} 
        &=E_{\bm{X}}\left[ E_{\bm{\theta}} \left[\int_0^\tau S_{\bm{\theta}}(t \; | \;A=1, \bm{X}_i=\bm{x}_i) - S_{\bm{\theta}}(t \; | \;A=0, \bm{X}_i=\bm{x}_i) \; dt\right ]\right ] \\
        &= \frac{1}{N} \sum_{i=1}^N 
        E_{\bm{\theta}} \left[\int_0^\tau S_{\bm{\theta}}(t \; | \;A=1, \bm{X}_i=\bm{x}_i) - S_{\bm{\theta}}(t \; | \;A=0, \bm{X}_i=\bm{x}_i) \; dt\right ] 
         \\ &= \frac{1}{N} \sum_{i=1}^N E_{\bm{\theta}}\Bigg[\sum_{h=1}^L \Big[ \pi_h\big(\bm{\psi}_{A1,1}(\bm{w}_i)^\prime\bm{\alpha}_h\big) \int_0^\tau \int_t^\infty K\big(s \; | \; \bm{\psi}_{A1,2}(\bm{w}_i)^\prime\bm{\beta}_h,\bm{\omega}_h\big) \; ds dt \\ & - \pi_h\big(\bm{\psi}_{A0,1}(\bm{w}_i)^\prime\bm{\alpha}_h\big) \int_0^\tau \int_t^\infty K\big(s \; | \; \bm{\psi}_{A0,2}(\bm{w}_i)^\prime\bm{\beta}_h,\bm{\omega}_h\big) \; ds dt \Big]\Bigg]
    \end{split}
\end{align}

\bibliographystyle{apalike}
\bibliography{refs.bib}

\end{document}